\documentclass[longauth]{aa}    
\usepackage{graphicx}
\usepackage{txfonts}
\usepackage{xspace}
\usepackage{hyperref}
\hypersetup{colorlinks=true, citecolor=blue, linkcolor=blue}

\newcommand{\tar}{J0148-4214}   
\newcommand{\bt}{BlackTHUNDER}  

\newcommand{\lmstar}{$\log(M_\star/M_\odot)=9.1$}

\newcommand{\civll}{C~{\sc{iv}}$_{1546,1551}$}
\newcommand{\civ}{C~{\sc{iv}}}
\newcommand{\heiiuv}{He~{\sc{ii}}$_{1640}$}
\newcommand{\ciiil}{C~{\sc{iii}]}$_{1909}$}
\newcommand{\ciii}{C~{\sc{iii}]}}
\newcommand{\oiill}{[O~{\sc{ii}}]$_{3726,3729}$}
\newcommand{\nii}{[N~{\sc{ii}}]}
\newcommand{\oiii}{[O~{\sc{iii}}]}
\newcommand{\oiiil}{[O~{\sc{iii}}]$_{5007}$}
\newcommand{\oiiill}{[O~{\sc{iii}}]$_{4959,5007}$}
\newcommand{\niill}{[N~{\sc{ii}}]$_{6548,6583}$}
\newcommand{\niil}{[N~{\sc{ii}}]$_{6583}$}
\newcommand{\ha}{H$\alpha$}
\newcommand{\hb}{H$\beta$}
\newcommand{\hg}{H$\gamma$}

\newcommand{\oiiia}{[O~{\sc{iii}}]$_{4363}$}
\newcommand{\siid}{[S~{\sc{ii}}]$_{6717,6731}$}

\newcommand{\oi}{O~{\sc{i}}$_{8446}$}
\newcommand{\oiop}{[O~{\sc{i}]}$_{6300}$}
\newcommand{\neiiib}{[Ne~{\sc{iii}]}$_{3869}$}
\newcommand{\neiiir}{[Ne~{\sc{iii}}]$_{3967}$}
\newcommand{\micron}{\ensuremath{\mu\mathrm{m}}\xspace}

\usepackage{newtxtext,newtxmath}

\begin{document} 

   \title{BlackTHUNDER: Evidence of three massive black holes in a $z\sim5$ galaxy}

   \author{Hannah {\"U}bler \inst{\ref{MPE}}\thanks{hannah@mpe.mpg.de} \and 
   Giovanni Mazzolari\inst{\ref{MPE}} \and
   Roberto Maiolino\inst{\ref{Kavli},\ref{Cavendish},\ref{UCL}} \and
   Francesco D'Eugenio\inst{\ref{Kavli},\ref{Cavendish}} \and 
   Nazanin Davari\inst{\ref{INAF_Rome}} \and   
   Ignas Juodžbalis\inst{\ref{Kavli},\ref{Cavendish}} \and 
   Michele Perna\inst{\ref{CAB}} \and
   Jorryt Matthee\inst{\ref{ISTA}} \and 
   Raffaella Schneider\inst{\ref{UniRome}} \and
   Rosa Valiante\inst{\ref{INAF_Rome}} \and 
   Santiago Arribas\inst{\ref{CAB}} \and
   Elena Bertola\inst{\ref{INAF_Arcetri}} \and 
   Andrew J.~Bunker\inst{\ref{Oxford}} \and
   Volker Bromm\inst{\ref{Texas}} \and 
   Stefano Carniani\inst{\ref{SNS}} \and 
   Stéphane Charlot\inst{\ref{Paris}} \and 
   Giovanni Cresci\inst{\ref{INAF_Arcetri}} \and
   Mirko Curti\inst{\ref{ESO}} \and 
   Richard Davies\inst{\ref{MPE}}  \and
   Frank Eisenhauer\inst{\ref{MPE},\ref{TUM}} \and
   Andrew Fabian\inst{\ref{IoA}} \and 
   Natascha M.~F\"orster Schreiber\inst{\ref{MPE}} \and 
   Reinhard Genzel\inst{\ref{MPE},\ref{UCB}} \and 
   Kohei Inayoshi\inst{\ref{KIAA}} \and
   Lucy R.~Ivey\inst{\ref{Kavli},\ref{Cavendish}} \and
   Gareth C. Jones\inst{\ref{Kavli},\ref{Cavendish}} \and
   Boyuan Liu\inst{\ref{ITA}} \and 
   Dieter Lutz\inst{\ref{MPE}} \and
   Daichi Kashino\inst{\ref{NAOJ}} \and
   Ruari Mackenzie\inst{\ref{ETH}, \ref{EPFL}} \and 
   Eleonora Parlanti\inst{\ref{SNS}} \and
   Brant Robertson\inst{\ref{SantaCruz}} \and 
   Bruno Rodríguez del Pino\inst{\ref{CAB}} \and
   T.~Taro Shimizu\inst{\ref{MPE}} \and
   Debora Sijacki\inst{\ref{Kavli}, \ref{IoA}} \and
   Eckhard Sturm\inst{\ref{MPE}} \and
   Sandro Tacchella\inst{\ref{Kavli},\ref{Cavendish}} \and 
   Linda Tacconi\inst{\ref{MPE}} \and
   Giulia Tozzi\inst{\ref{MPE}} \and
   Alessandro Trinca\inst{\ref{Como},\ref{INAF_Rome}} \and 
   Giacomo Venturi\inst{\ref{SNS}} \and 
   Marta Volonteri\inst{\ref{Paris}} \and 
   Chris Willott\inst{\ref{Herzberg}} \and 
   Saiyang Zhang\inst{\ref{TexasPhy}, \ref{Weinberg}}
          }

   \institute{
   Max-Planck-Institut f\"ur extraterrestrische Physik, Gie{\ss}enbachstra{\ss}e 1, 85748 Garching, Germany\label{MPE}  \and
    Kavli Institute for Cosmology, University of Cambridge, Madingley Road, Cambridge, CB3 0HA, UK\label{Kavli} \and
    Cavendish Laboratory, University of Cambridge, 19 JJ Thomson Avenue, Cambridge, CB3 0HE, UK\label{Cavendish} \and
    Department of Physics and Astronomy, University College London, Gower Street, London WC1E 6BT, UK\label{UCL} \and
    INAF/Osservatorio Astronomico di Roma, Via Frascati 33, 00040 Monte Porzio Catone, Italy\label{INAF_Rome} \and
    Centro de Astrobiolog\'{\i}a (CAB), CSIC-INTA, Ctra. de Ajalvir km 4, Torrej\'on de Ardoz, E-28850, Madrid, Spain\label{CAB} \and 
    Institute of Science and Technology Austria (ISTA), Am Campus 1, 3400 Klosterneuburg, Austria\label{ISTA} \and 
    Dipartimento di Fisica, Sapienza UniversitÃ  di Roma, P.le A. Moro 2, 00185 Roma, Italy\label{UniRome} \and 
    INAF - Osservatorio Astrofisco di Arcetri, largo E. Fermi 5, 50127 Firenze, Italy\label{INAF_Arcetri} \and 
    Department of Physics, University of Oxford, Denys Wilkinson Building, Keble Road, Oxford OX13RH, UK\label{Oxford} \and 
    Department of Astronomy, University of Texas, Austin, TX 78712, USA\label{Texas} \and
    Scuola Normale Superiore, Piazza dei Cavalieri 7, I-56126 Pisa, Italy\label{SNS} \and
    Sorbonne Universit\'e, CNRS, UMR 7095, Institut d'Astrophysique de Paris, 98 bis bd Arago, 75014 Paris, France\label{Paris} \and
    European Southern Observatory, Karl-Schwarzschild-Strasse 2, 85748 Garching, Germany\label{ESO} \and
    Department of Physics, TUM School of Natural Sciences, Technical University of Munich, 85748 Garching, Germany\label{TUM} \and
    Institute of Astronomy, Madingley Road, Cambridge CB3 0HA, UK\label{IoA} \and    
    Departments of Physics and Astronomy, University of California, Berkeley, CA 94720, USA\label{UCB} \and
    Kavli Institute for Astronomy and Astrophysics, Peking University, Beijing 100871, China\label{KIAA} \and
    Universit\"at Heidelberg, Zentrum f\"ur Astronomie, Institut f\"ur Theoretische Astrophysik, Albert Ueberle Str. 2, D-69120 Heidelberg, Germany\label{ITA} \and
    National Astronomical Observatory of Japan (NAOJ), 2-21-1, Osawa, Mitaka, Tokyo 181-8588, Japan\label{NAOJ} \and
    Department of Physics, ETH Z{\"u}rich, Wolfgang-Pauli-Strasse 27, Z{\"u}rich, 8093, Switzerland\label{ETH} \and
    Institute of Physics, Laboratory of Astrophysics, Ecole Polytechnique F\'ed\'erale de Lausanne (EPFL), Observatoire de Sauverny, 1290 Versoix, Switzerland\label{EPFL} \and
    Department of Astronomy and Astrophysics, University of California, Santa Cruz, 1156 High Street, Santa Cruz, CA 95064 USA\label{SantaCruz} \and 
    Como Lake Center for Astrophysiscs, DiSAT Universit$\`a$ degli Studi dell'Insubria, via Valleggio 11, 22100, Como, Italy\label{Como} \and 
    NRC Herzberg, 5071 West Saanich Rd, Victoria, BC V9E 2E7, Canada\label{Herzberg} \and
    Department of Physics, University of Texas at Austin, Austin, TX 78712, USA\label{TexasPhy} \and
    Weinberg Institute for Theoretical Physics, Texas Center for Cosmology and Astroparticle Physics, University of Texas at Austin, Austin, TX 78712, USA\label{Weinberg} 
    }

  \abstract 
  {We present observational evidence of three massive, accreting black holes in the $z=5.0167$ galaxy J0148-4214 from JWST/NIRSpec-IFU spectroscopy. The black holes are revealed through broad H$\alpha$ emission (FWHM = 430--2920 km/s) without a forbidden-line counterpart in the bright [O~{\sc{iii}}] doublet. Channel maps of the asymmetric central H$\alpha$ profile isolate two spatially distinct broad-line regions (BLRs), separated by $190\pm40$~pc, while a third BLR is found in the galaxy outskirts with a projected separation of 1.7 kpc. 
  We discuss whether this emission could be due to supernovae, shocks, winds, or massive stars, but find the BLR origin most likely.
  Using single-epoch virial relations, we estimated black hole masses of $\log(M_\bullet/M_\odot)=7.9\pm0.4$ (primary), $5.8\pm0.5$ (secondary), and $6.3\pm0.5$ (third off-nuclear). We argue that the two central black holes will likely rapidly merge, with a simple dynamical friction time estimate of the order of $\lesssim700$~Myr. Assuming that   the third off-nuclear black hole is also in the process of sinking towards the centre, it will likely lead to a second merger, and we investigated the detection probability of such mergers with LISA. Alternatively, the third black hole may be the result of a previous central three-body interaction or a gravitational recoil, where our observations would provide evidence that such black holes may retain their accretion discs and BLRs even in the aftermath of such extreme dynamical interactions. The possible discovery of a black hole triplet at high redshifts, together with other recent results on distant black hole pairs, indicates that multiple massive black hole systems were likely common in the early Universe. Our results highlight the importance of IFU observations for the detection of massive black hole multiplets in distant galaxies, the progenitors of massive black hole mergers that may be detected with next-generation gravitational wave observatories.}
  
   \keywords{galaxies: high-redshift -- galaxies: active -- galaxies: supermassive black holes}

   \maketitle

\section{Introduction}\label{s:intro}

There are essentially two ways of growing massive black holes following their formation: via gas accretion and through black hole mergers. The former is thought to be the dominant growth mode for the majority of massive black holes and across cosmic time \citep[e.g.][]{Soltan82, Volonteri03, Sijacki07}. However,  some models and simulations indicate that for certain black hole mass ranges and cosmic epochs, mergers could contribute significantly to massive black hole growth \citep[e.g.][]{Dubois14, Kulier15, Pacucci20, Piana21, Bhowmick25}.

Mergers of massive black holes have been predicted as a consequence of galaxy collisions \citep[e.g.][]{Begelman80, Kauffmann00, Menou01, Volonteri03, diMatteo05, Ferrarese05, Mayer07, Khan16, Rantala17,  Tremmel17, Volonteri20, Mannerkoski22}, and recent observations with JWST indeed confirm that the major merger rate at $z>3$ is significant, on the order of $3{-}8$ galaxies per Gyr, and even higher when accounting for minor mergers \citep[e.g.][]{Dalmasso24, Puskas25, Duan25}.
Observationally, massive black hole pairs on sub-kiloparsec separations to separations of a few tens of kiloparsecs  have been identified across a wide redshift range (e.g.\ \citealp{Junkkarinen01, Komossa03, Comerford09, Comerford13, Comerford15, Green10, Husemann18, Mannucci22, Koss23, Neureiter23, Perna23, Perna25a, Scialpi24, Uebler24b, Matsuoka24, Zamora24, Ulivi25}; see also \citealp{Tanaka24, Merida25}), including some triplets in the nearby Universe \citep{Deane14, LiuX19, Pfeifle19, Kollatschny20}. 
The identification of multi-black hole systems is aided by their emission signatures during an active galactic nucleus (AGN) phase. 
There is strong theoretical evidence that black holes also grow through enhanced gas accretion during the galaxy merging process (e.g.\ \citealp{Kormendy01, diMatteo05, Springel05b, Hopkins08, Johansson09, vanWassenhove12, Capelo15, Liao17, Blumenthal18, Saeedzadeh24, Talbot24}; but see also \citealp{Steinborn16}). 

During these phases of active growth, massive black holes can be detected through various spectral signatures, including the broad, permitted emission of gas rotating at high velocities ($\geq$1000~km/s) around the black hole, forming the broad-line region (BLR); characteristic narrow-line ratios indicative of the hard ionizing radiation from an AGN; $X$-ray emission; or high-velocity ($\geq$1000~km/s) outflows in both permitted and forbidden lines \citep[e.g.][and references therein]{Padovani17}. Among these signatures, emission from the BLR is of particular interest because it allows  a measurement of the black hole mass, assuming that the gas in the BLR is in virial equilibrium, or at least that its width and luminosity scale with black hole mass as calibrated locally \citep[e.g.][]{Kaspi00, Greene05}.

Broad-line regions of luminous quasars have been detected out to $z$$\sim$$7$ and used to estimate black hole masses, mainly derived from Mg~{\sc{ii}}$_{2796,2803}$ or C~IV$_{1548,1551}$ \citep[e.g.][]{Willott03, Willott10, Jiang07, Kurk07, Trakhtenbrot17}. 
 However, only in the last few years, BLR emission from bright hydrogen and helium lines has become accessible in $z$$\gtrsim$$4$ AGNs and quasars, thanks to the spectroscopic capabilities of JWST in the infrared regime. Since then, a large number of high-$z$ broad-line (Type~I) AGNs have been discovered \citep[e.g.][]{Kocevski23, Uebler23, Harikane23, Greene23, Kokorev23, Matthee24, Maiolino24,  Furtak24, Juodzbalis26, Taylor25, Tripodi25}. The detection of an  off-centre broad \hb\ line also allowed for the discovery of a close-separation massive black hole pair in a galaxy merger at $z$$=$$7.15$, with a projected separation of the two nuclei of only 620~pc \citep{Uebler24b}. However, in this merger system the second black hole is a narrow-line (Type II) AGN, and only the mass of the off-centre black hole could be measured. 

In this paper we present observations of the galaxy \tar\ at $z\sim5.017$ obtained with the Integral Field Spectroscopy (IFS) mode of the Near-Infrared Spectrograph (NIRSpec) on board JWST \citep{Jakobsen22, Boeker22}. We report the discovery of multiple broad \ha\ emission line regions in \tar, providing evidence of three massive accreting black holes within $0.27\arcsec$ (projected distance of 1.7~kpc). 
We describe the available data and processing of the NIRSpec-IFS data in Sect.~\ref{s:data}.
The discovery of multiple broad emission line regions and their possible interpretations are discussed in Sect.~\ref{s:analysis}. Global mass estimates for \tar\ are described in Sect.~\ref{s:global}.
In Sect.~\ref{s:bhprops} we discuss the properties of the black holes in \tar, their relation to their host galaxy, and their likely fate. 
We conclude in Sect.~\ref{s:conclusions}.

Throughout this work, we adopt a \cite{Chabrier03} initial mass function ($0.1-100 M_\odot$) and a flat $\Lambda$CDM cosmology with $H_0=70$~km~s$^{-1}$~Mpc$^{-1}$, $\Omega_\Lambda=0.7$, and $\Omega_m=0.3$.

\section{Observations and data processing}\label{s:data}

\subsection{JWST/NIRCam data}\label{s:nircam}
\begin{figure*}
    \centering
        \includegraphics[width=0.95\textwidth]{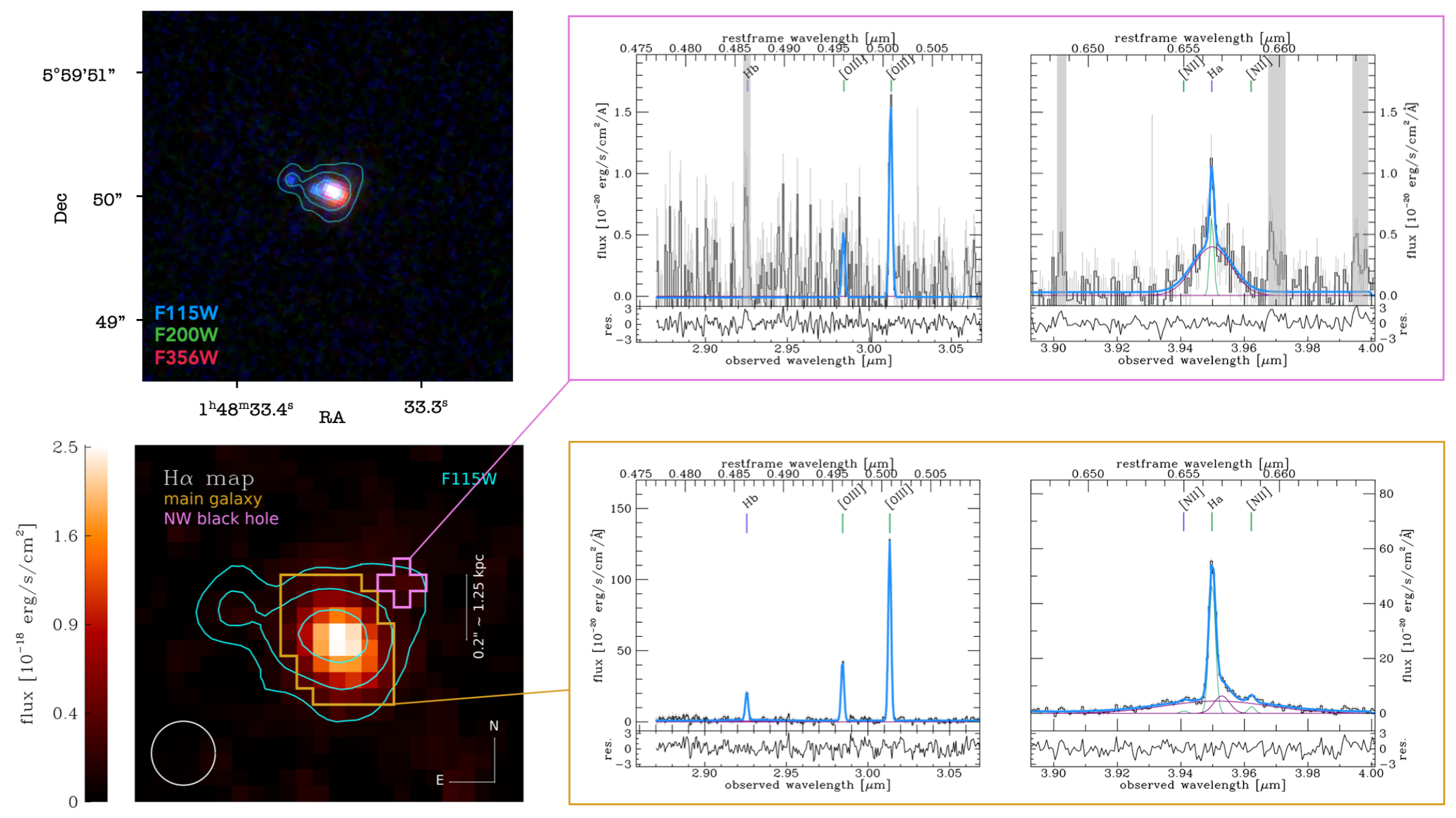}
    \caption{\textit{Top left}: RGB NIRCam image of \tar\ \citep[see also][]{Matthee24}. The blue contours are drawn at 0.009, 0.021, and 0.057 MJy/sr on the F115W image. The outermost contour encompasses the main target, the two smaller blue sources, and faint emission to the north-west of \tar, where we locate a Type~I AGN (see Sect.~\ref{s:blrs}). The cutout size is $3\arcsec\times3\arcsec$.
    \textit{Bottom left}: H$\alpha$ line map integrated in the observed wavelength range $\lambda=3.9-4.0\mu$m. The white circle shows the approximate PSF FWHM at the wavelength of \ha. The cyan contours are the same as in the upper panel. Spectra extracted from the regions outlined in pink (NW black hole) and orange (main galaxy) are shown in the \textit{top} and \textit{bottom right} panels, respectively. We mark several artefacts with grey shading. Fits to the spectra are shown in blue, while the narrow and broad components are shown in green and purple, respectively. Residuals are plotted as (data-model)/error. While the permitted \oiii\ lines show no evidence of a broad component in the emission lines profile, \ha\ displays a complex profile for the main galaxy (bottom right panel), requiring two separate broad components (FWHM of $2920\pm140$ and $430\pm80$~km/s). In Sect.~\ref{s:blrs} we argue that this profile indicates the presence of two AGNs in the main galaxy. The broad H$\alpha$ profile seen to the north-west of the main galaxy indicates the presence of a third AGN (\textit{top right}).}
    \label{f:overview}
\end{figure*}

\tar\ (R.A. 01:48:33.29, Dec. +05:59:50.04; J2000) was observed as part of the EIGER GTO survey (PID: 1243; PI: S.~Lilly) in JWST/NIRCam WFSS mode with the F356W filter
($\sim3.1-4.0$\micron; spectral resolution $R\sim1600$) and simultaneous imaging in F115W, F200W, and F356W. The data reduction is described by \cite{Kashino23}. \tar\ was identified as a $z=5.019$ Type~I AGN by \cite{Matthee24}. Using the single-epoch calibration by \cite{Reines13}, they found a single black hole with mass $\log(M_\bullet/M_\odot)_{\rm H\alpha}=7.32\pm0.10$ (excluding systematic uncertainties and dust correction) and bolometric luminosity $L_{\rm bol}=(5.9\pm0.4)\times10^{44}$~erg/s. The imaging data presented by \cite{Matthee24} show a bright, slightly elongated red source and two small fainter and bluer sources within $0.5\arcsec$ in projection. In the top left panel of Fig.~\ref{f:overview} we show a $3\arcsec\times 3\arcsec$ NIRCam false-colour composite of \tar. F115W contours encompass the three distinct sources; faint emission to the north-west of the main target is also seen.

\subsection{JWST/NIRSpec-IFU data}\label{s:ifs}

\tar\ was subsequently observed in JWST/NIRSpec-IFU mode as part of the \bt\ survey (PID: 5015; PIs: H.~\"Ubler, R.~Maiolino). The NIRSpec data were taken on December 15, 2024, with a medium cycling pattern of eight dither positions and a total integration time of about 1~h with the high-resolution grating/filter pair G395H/F290LP, covering the wavelength range $2.87-5.14\mu$m ($R\sim2000-3500$), and about 0.5~h with PRISM/CLEAR ($\lambda=0.6-5.3\mu$m; $R\sim30-300$).

Raw data files were downloaded from the Barbara A.~Mikulski Archive for Space Telescopes (MAST) and subsequently processed with the JWST Science Calibration pipeline\footnote{\scriptsize \url{https://jwst-pipeline.readthedocs.io/en/stable/}} version 1.15.0 under the Calibration Reference Data System (CRDS) context jwst\_1293.pmap. We made several modifications to the default reduction steps to increase data quality, which are described in detail by \cite{Perna23} and which we briefly summarise here. 
Count-rate frames were corrected for $1/f$ noise through a polynomial fit. 
During calibration in Stage 2, we removed regions affected by failed open MSA shutters, cosmic rays and bad pixels.
Remaining outliers were flagged in individual exposures following \cite{DEugenio23}; we rejected the 98\textsuperscript{th} (99.5\textsuperscript{th}) percentile of the resulting distribution in the grating (prism) observations.
The final cubes were combined using the `drizzle' method with a pixel scale of $0.05''$.

No target acquisition was included in our NIRSpec observing setup. To correct for the resulting astrometric inaccuracies, we use the NIRSpec-IFS prism data to construct pseudo images matching the available NIRCam data. Registering our pseudo images to the NIRCam images (which are in turn registered to {\it Gaia} Data release 2; \citealp{Gaia2_2018}; see \citealp{Kashino23}), we find an average offset of about $0.21\arcsec$. This is comparable to offsets found for other NIRSpec-IFS cubes \citep[e.g.][]{Arribas24, Jones24a, Jones24b, Lamperti24, Uebler24c, Parlanti25}.

We use spaxels away from the central source and free of line emission to perform a background subtraction.
To obtain realistic flux uncertainties, for each spectral extraction performed during our analysis, we scale the formal error extension of the final data cube to match the rms in regions free of spectral features \citep[see][]{Uebler23}.
From simultaneous fitting of multiple emission lines in the NIRSpec-IFS high-resolution data, we measure a spectroscopic redshift of $z=5.01672\pm0.00002$ for \tar.

\section{Discovery of multiple broad emission line regions}\label{s:analysis}

The high-resolution cube of \tar\ shows several strong rest-frame optical emission lines: \hb, \oiiill, \oiop, \ha\ and \niill\ (\siid\ is seen but falls partly into the detector gap and is very noisy). In the bottom left panel of Fig.~\ref{f:overview} we show a line map integrated in the wavelength range 3.9-4.0\micron, covering \ha\ and any \nii\ emission, zoomed in on the position of \tar. The cyan contours derived from the NIRCam F115W image are the same as in the top left panel. 
We show the approximate point-spread function (PSF) FWHM at the wavelength of \ha, which is $\sim0.197^{\prime\prime}$, as a white circle \citep[see][]{JonesG26}. We indicate two extraction regions in orange and pink. The orange region includes emission from the main target; the extraction aperture is slightly stretched along the NE-SW direction, following the shape of the \ha\ emission. We note that this is also the orientation of the slicers in our NIRSpec-IFS observations, and the PSF may be elongated in slicer direction \citep[see][]{DEugenio23}. The corresponding integrated spectra around \oiii+\hb\ and \ha+\nii\ are shown in the bottom right panels. We show the full, high-resolution spectrum extracted from the main galaxy aperture in Appendix~\ref{a:fullspec}. The \ha\ emission line exhibits a broad, asymmetric profile, possibly due to a superposition of two broad components. No corresponding broad emission is seen at the wavelength position of the brighter \oiiil\ line.
In the top right panels we show the integrated spectra extracted from the pink region,\footnote{We note that the conclusions presented in this paper remain unchanged if we increase the diameter of the pink extraction region to 5 pixels ($\sim1.3\times{\rm FWHM_{PSF}}$).}
offset by about 0.27\arcsec (1.7~kpc) from the main source. Another broad component symmetric around \ha\ is clearly detected. Again, no corresponding broad component is detected in [O~{\sc{iii}}]$_{5007}$. 
We perform a systematic search for additional broad \ha\ lines around the main galaxy aperture, using the pink extraction region as a template. While narrow \oiiil\ and at lower significance narrow \ha\ are detected for all test apertures, we find no evidence of further broad \ha\ components.

In the following paragraphs, we discuss different physical scenarios that could give rise to the broad \ha\ components detected in \tar. We begin with our favoured interpretation, three accreting massive black holes, and discuss and discard various other scenarios.

\subsection{Broad-line regions from three massive black holes}\label{s:blrs}

Broad emission line profiles in permitted lines without a counterpart in forbidden lines are commonly interpreted as emission from the broad-line region of accreting massive black holes. In \tar\, we find evidence of three such BLRs.
First, the broad \ha\ profile extracted from the north-western (pink) region and detached from the broad \ha\ emission in the centre is clearly distinct in shape to the central emission: it is less broad than the central broad \ha\ emission and symmetric around the narrow \ha\ component. 
The broad \ha\ component has no counterpart in forbidden lines, specifically not in the \oiiill\ doublet. Therefore, it is unlikely to be associated with high-velocity gas in the interstellar medium (see Sect.~\ref{a:alternatives}). Instead,
we here interpret this emission as the BLR of a black hole located in the outskirts of \tar. At this off-nuclear position, it may be classified as an offset AGN \citep{Comerford14}, either in the process of being accreted into the centre of \tar, or possibly having been ejected in a previous multi-body interaction or through a recoil kick of a previous black hole merger \citep{Peres62, Bekenstein73, Saslaw74, Mikkola90, Volonteri03, Madau04, Merritt04, Sijacki11, Blecha16, Mannerkoski22, Partmann24}.

Second, the broad \ha\ emission extracted from the central region of \tar\ (orange region) is clearly asymmetric. Similar profiles have been found for two AGNs at $z=4-6$ observed with NIRSpec-MSA in the JADES survey \citep{Eisenstein23} by \cite{Maiolino24}, who interpreted them as signatures of two AGNs within the MSA slit ($\lesssim0.2\arcsec$) through double-Gaussian fits. Without further spatial information such an interpretation remains uncertain, because asymmetric BLRs can also be caused by an anisotropic gas distribution, inflows or outflows in the BLR, or a specific BLR geometry \citep[e.g.][and references therein]{StorchiBergmann17}. However, if there are indeed two AGNs, the spatial information afforded through NIRSpec-IFS can help to separate their emission features \cite[see][]{Uebler24b}.
While we have no evidence of two central emission peaks in the \ha\ line map derived from our NIRSpec-IFS data, a visual inspection of the central region of \tar\ indeed suggests changes of the broad-line profile shape across the central spaxels. 

\begin{figure}
    \centering  
    \includegraphics[width=0.85\columnwidth]{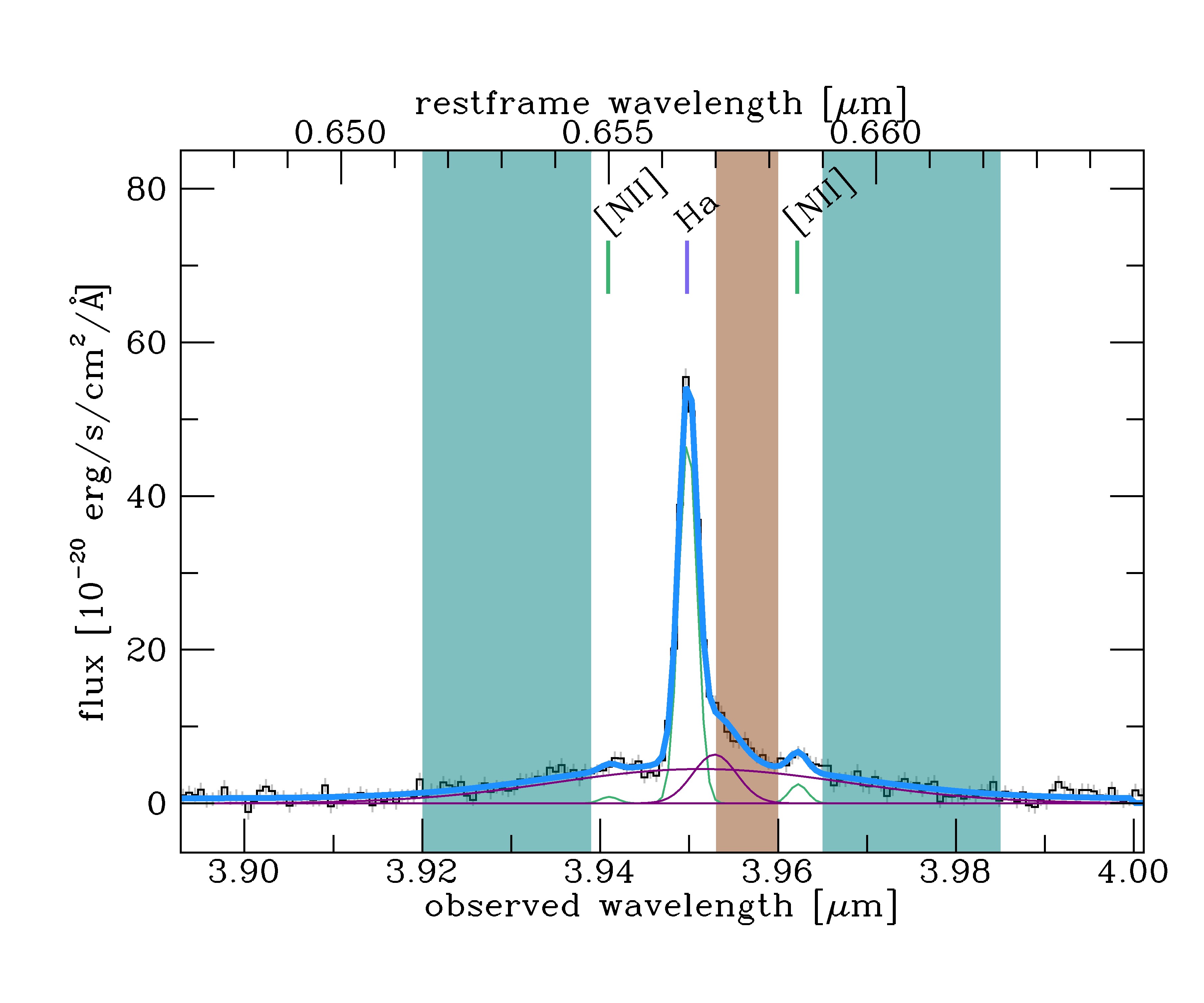} 
    \includegraphics[width=0.65\columnwidth]{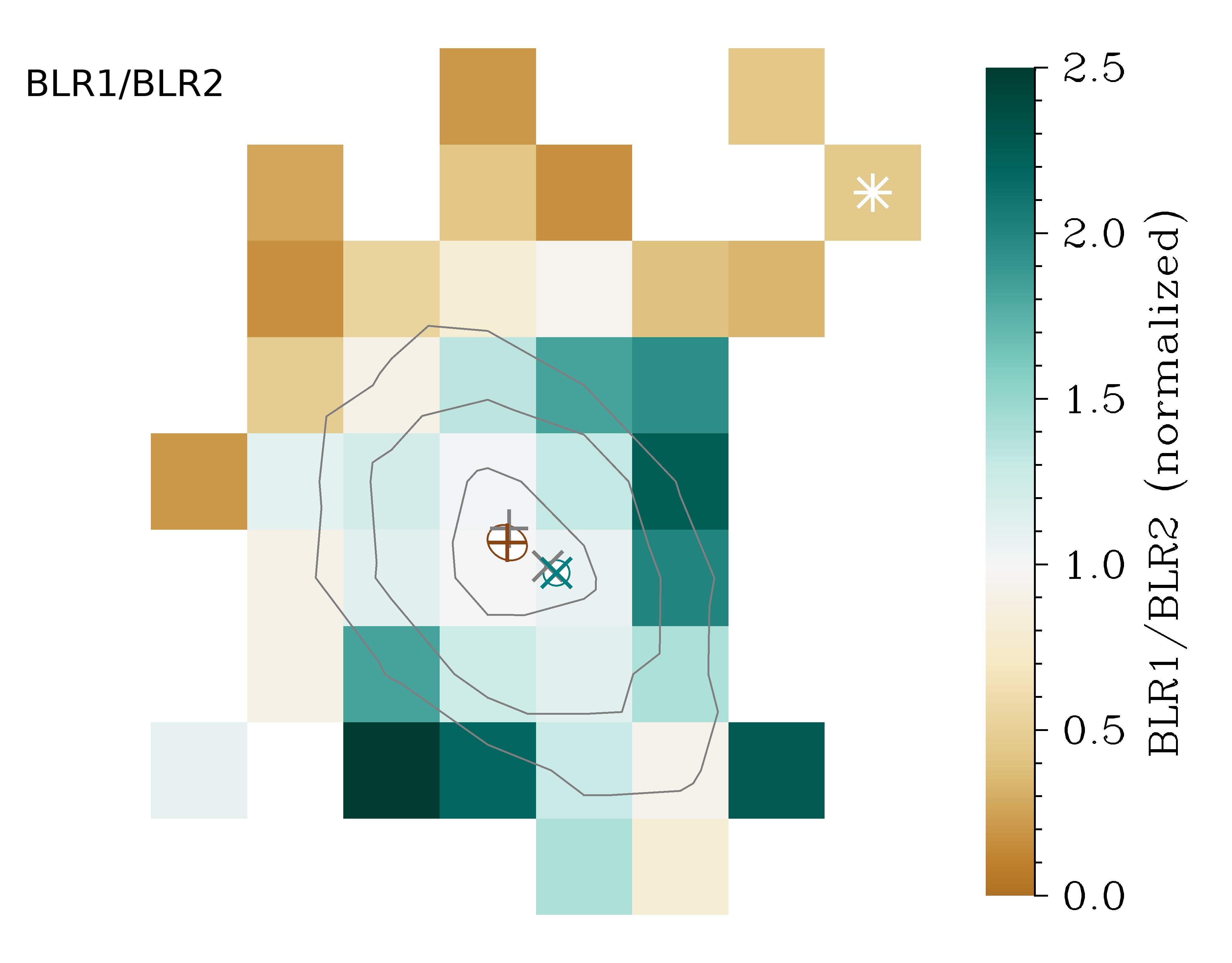}
    \caption{\textit{Top}: Illustration of the extraction regions for the BLR channel maps, as described in Sect.~\ref{s:blrs}. The wavelength ranges to construct BLR1 and BLR2 are shown in teal and brown shading, respectively. We note that in this illustration, BLR2 is contaminated by BLR1, and we subtracted the contamination as described in the main text to obtain a clean channel map for BLR2. 
    \textit{Bottom}: Map of the ratio BLR1/BLR2 after normalisation to peak flux. We plot pixels with $S/N\geq0.5$ and draw contours at $S/N=3, 5, 8$ for reference, where the $S/N$ is calculated from the normalised ratio BLR1/BLR2. The teal cross and brown plus show the centroids of the channel maps for BLR1 and BLR2, with ellipses indicating two times the centroid uncertainty, derived as described in Sect.~\ref{s:blrs}. As grey symbols we show the centroids derived from maps including only pixels with $S/N\geq2$ (uncertainties are omitted for clarity). The white star marks the position of the third BLR detected in \tar. A modulation from the north-eastern region to the south-western region is visible, as expected for the images of two point sources spatially separated along that axis.}
    \label{f:blrmaps}
\end{figure}

For a quantitative investigation we employ a spectro-astrometric approach \citep[e.g.][]{Beckers82, Bailey98, Gnerucci10} and constructed channel maps of the two putative BLRs. Here, it is important to avoid any narrow line emission from \ha\ or \niill\ which would contaminate the BLR light maps. Therefore, we proceed as follows: after subtraction of the (faint) continuum emission we derived a map of the broader BLR component (BLR1) by integrating emission at observed wavelengths $\lambda_{\rm obs}=3.92-3.939\mu$m and $\lambda_{\rm obs}=3.965-3.985\mu$m, i.e.~omitting any narrow \ha\ and \niill\ emission from the host galaxy, as well as the putative second BLR component (see teal shading in the top panel of Fig.~\ref{f:blrmaps}). Next, we derived a map of the narrower BLR component (BLR2), contaminated by BLR1, by integrating emission at observed wavelengths $\lambda_{\rm obs}=3.953-3.96\mu$m (brown shading in top panel of Fig.~\ref{f:blrmaps}), i.e.~again omitting any narrow \ha\ and \niill\ emission from the host galaxy. 
We used our best fit to the central aperture spectrum (see below) to calculate the fractional contribution of flux from BLR1 to the total flux in this wavelength range with respect to the flux in our BLR1 map. We used the BLR1 map to subtract this fractional contribution from the contaminated BLR2 map, to obtain a clean map of BLR2.
The resulting maps are shown in Fig.~\ref{f:channelmaps}. They have a distinct flux distribution, indicating that the asymmetric broad emission does not originate from a single point source. 
We determined the centroids of the two sources through two-dimensional Gaussian fits to the flux distributions of the two maps (see Appendix~\ref{a:channelmaps} for details on the precision). We used a box with half window size $\sim3.5\times{\rm FWHM_{PSF}}$. Uncertainties are derived through Monte Carlo methods by varying the input images 10000 times each with their associated noise, repeating the fits (see Appendix~\ref{a:channelmaps}). Then, we derived a projected spatial separation of $d_{\rm I,II}=190\pm40$~pc between the centroids of BLR1 and BLR2 ($\sim0.03$\arcsec), where the uncertainty on the measurement calculated from the best-fit uncertainty is consistent with the uncertainty derived from the Monte Carlo runs. 
As an additional test, we measured the centroids including only pixels with $S/N\geq2$ (uncertainties are again derived through 10000 Monte Carlo runs). The such derived centroids agree with our above measurements within the uncertainties. The corresponding projected spatial separation of BLR1 and BLR2 from this method is $d_{\rm I,II}=160^{+40}_{-30}$~pc ($\sim0.026$\arcsec), consistent with our first measurement.
In Appendix~\ref{a:blrtests} we describe a mock test for the recovery of the distance between the BLRs from our channel maps. This gives consistent results ($d_{\rm I,II, mock}=200^{+60}_{-40}$~pc), though indicates that the uncertainty associated with our distance measurement may be underestimated by $10-20$ pc.
We note that the offset we find is much larger than the BLR size associated with a single black hole \citep[typically sub-parsec to $\sim$parsec scales; e.g.][]{Kaspi00, Bentz13, Santos24, Davies25}. 

To further highlight the significance of the offset between the two BLRs, in the bottom panel of Fig.~\ref{f:blrmaps} we show the ratio of the two BLR maps (after normalisation to peak flux), which illustrates a NE-SW pattern, as expected from two images of two point sources that are spatially offset along that axis (see also Appendix~\ref{a:blrtests}).
We emphasise that we would not expect to see any of these signatures if the asymmetric broad \ha\ emission originated from a single source, i.e.~a black hole with a peculiar BLR geometry, including   absorption or asymmetric emission or velocity distribution within the BLR. We reiterate that typical BLR sizes are on $\lesssim$~parsec scales, i.e.~at least 150 times smaller than the centroid separation we measure.

For further analysis, we performed a multi-component Gaussian fitting of the high-resolution spectra, assuming that the broad emission components stem from three massive, accreting black holes.
We included a narrow Gaussian for \hb, \oiiill, \oiop, \ha, \niill. We do not fit \siid\ because the lines are close to the detector gap in a region of elevated noise. The intensity ratio of the \oiii\ doublet is fixed to [O~{\sc{iii}}]$_{\lambda5007}$/[O~{\sc{iii}}]$_{\lambda4959}=2.98$, as given by the Einstein coefficients of the two levels; similarly, the intensity ratio of the [N~{\sc{ii}}] doublet is fixed to [N~{\sc{ii}}]$_{\lambda6584}$/[N~{\sc{ii}}]$_{\lambda6548}=2.94$ \citep[e.g.][]{Osterbrock06}. The narrow line emission is assumed to trace the interstellar medium in \tar. We tied the line centroids and line widths separately for [O~{\sc{iii}}]+H$\beta$ and for H$\alpha$+[N~{\sc{ii}}]. These groups of lines are close enough in wavelength to not expect significant differences in the instrument line-spread function \citep[LSF; see][]{Jakobsen22}. 
We also included a linear polynomial to account for any continuum emission.
To model the BLRs in \tar, we included two broad Gaussians for \ha\ and \hb\ for the spectrum extracted from the central region (Fig.~\ref{f:overview}, orange aperture), and one broad Gaussian for the spectrum from the north-west (pink aperture). In both cases, the relative velocities and line widths are tied between \ha\ and \hb.
The best-fit emission line fluxes are listed in Table~\ref{t:fluxes_fiducial}.
We highlight that the integrated flux ratio of the two broad \ha\ components in the centre is about 0.2. This is different from results for double-Gaussian fits which have been performed for some high-$z$ AGNs with symmetric broad components, which typically yield values around 0.5 and indicate a non-Gaussian shape of the BLR \citep{DEugenio25a, DEugenio25b}.
In general, non-Gaussian BLR shapes have been observed for a few high $S/N$, high-$z$ AGNs \citep[e.g.][]{DEugenio25b, Rusakov25, Torralba25} and more frequently for AGNs in the nearby Universe (e.g.~\citealp{Kollatschny11, Kollatschny13}, and references therein; see also~\citealp{Ji25b}). For the case of \tar\ the Gaussian decomposition provides an adequate fit to the data, as evidenced by the small and regular residuals (Fig.~\ref{f:overview}, right panels).  However, it is possible that higher $S/N$ or higher spectral resolution data could reveal deviations from Gaussian shapes. A systematic investigation of the BLR shapes and possible parametrisations for the full BlackTHUNDER sample will be presented in a forthcoming work.

We note the velocity offsets of the three broad components with respect to the central narrow \ha\ emission from our fits. For the primary broad \ha\ component and the secondary broad \ha\ component we find velocity offsets of $\Delta v=110$ and $\Delta v=228$~km/s,  respectively, while the third, off-nuclear broad \ha\ component has a velocity offset of only $\Delta v=3$~km/s (or $\Delta v=10$~km/s with respect to the narrow \ha\ component at the off-nuclear position).

From our fiducial model, we find a narrow flux ratio \ha$_{\rm n}$/\hb$_{\rm n}=3.0\pm0.2$ for the main galaxy spectrum,  potentially indicating a low attenuation of $A_V\sim0.13$, using the attenuation law by \cite{Calzetti00} or the extinction curve by \cite{Gordon03}, but also consistent with no dust attenuation under Case~B recombination. For the BLR, values higher than the canonical Case~B are routinely observed \citep[e.g.][]{Osterbrock77, Osterbrock81, Crenshaw88, LaMura07, Ilic12}. A broad component in \hb\ corresponding to the primary AGN detected from the \ha\ emission is preferred by the fit, although it is only marginally detected. We find a flux ratio of \ha$_{\rm BLR1}$/\hb$_{\rm BLR1}\sim11$. If this ratio is (partly) due to reddening, instead of collisional excitation due to the much higher densities in the BLR, our black hole mass estimates (see Sect.~\ref{s:mbh}) would correspond to lower limits.
We followed a similar approach to fit the low-resolution prism observations, described in Appendix~\ref{a:prism}.

\begin{table}
\small
\caption{$R$$\sim$$2700$ emission line fluxes.}\label{t:fluxes_fiducial}%
\begin{tabular}{@{}lc@{}}
 & Main Galaxy \\
\hline
H$\beta_{\rm n}$ & $4.28\pm0.25$  \\
H$\beta_{\rm BLR~1}$ & $1.69\pm0.90$  \\
\oiiil & $28.33\pm0.38$ \\
\oiop\ & $0.59\pm0.12$  \\
H$\alpha_{\rm n}$ & $12.14\pm0.59$  \\
H$\alpha_{\rm BLR~1}$ & $18.27\pm1.44$  \\
H$\alpha_{\rm BLR~2}$ & $3.91\pm0.86$ \\
\niil & $0.63\pm0.14$  \\
\hline
& North-western region \\
\oiiil &  $3.83\pm0.70$ \\
H$\alpha_{\rm n}$ & $1.07\pm0.45$ \\
H$\alpha_{\rm BLR~3}$ & $5.98\pm1.41$ \\
\end{tabular}
\tablefoot{ 
Observed $R$$\sim$$2700$ emission line fluxes in units of $10^{-18}$~erg/s/cm$^2$, extracted assuming our fiducial model: two central BLRs (orange aperture in Fig.~\ref{f:overview}; main galaxy) and one BLR in the north-western region (pink aperture in Fig.~\ref{f:overview}; NW black hole).  We note that no second component is preferred by the fit for the much fainter broad \hb\ emission in the centre, and also that the primary BLR component is only marginally detected in \hb. For the north-western region, we do not report fluxes for \hb\ because of an artefact at this line position; \oiop\ and \nii\ are not significantly detected.
}
\end{table}

For the high-resolution data, we also performed a spaxel-by-spaxel fitting for the main galaxy to constrain the narrow-line kinematics. To this end, we fixed the line centroids and widths of the BLR components to the best-fit values from the main galaxy aperture spectrum (Fig.~\ref{f:overview}), and otherwise proceeded as for our fiducial fit as described above. Resulting flux, velocity and velocity dispersion maps for \oiiil\ are shown in Appendix~\ref{a:narrowmaps}.

\subsection{A single black hole in the centre}\label{a:alternatives}

We explore alternative interpretations of the broad \ha\ emission from the main galaxy aperture.
In the discovery paper by \cite{Matthee24} the authors consider a fit with a single BLR component for \ha\ (in addition to narrow \ha\ emission and broad \nii\ emission) to their lower spectral resolution NIRCam grism data, estimating a (single) black hole mass of $\log(M_\bullet/M_\odot)$$=$$7.32$$\pm$$0.10$. If we fit a single \ha\ BLR component in addition to the narrow line emission to the main galaxy aperture spectrum, we find $\log(M_\bullet/M_\odot)$$=$$7.69$$\pm$$0.42$, consistent within the uncertainties with the estimate by \cite{Matthee24}. Their somewhat lower black hole mass estimate can be explained through their choice of a broad \nii\ component, while we include only a narrow line contribution for \nii, informed by the line profile of other forbidden emission lines, especially \oiii, which was not available from the NIRCam grism observations.
However, the asymmetric profile of the broad \ha\ component, which is more pronounced in our high-resolution data, is not well fit by a single Gaussian, and we therefore disregard this model.

 However, there could be a single black hole with a non-Gaussian, asymmetric BLR shape, for instance due to strong inflows or outflows in the BLR \citep[e.g.][]{Rigamonti25, Rosborough25}, or due to other anisotropic gas distributions. Asymmetric BLR shapes are known in AGNs in the local Universe, including in long-term monitoring campaigns (see e.g.\ \citealp{Peterson02, Lu22} for galaxy NGC~5548; \citealp{Fries24}).
However, if the emission originates from within a single BLR (i.e.~a point source for the cosmological distance of \tar), we would have expected the same flux distribution from our mapping exercise of the two components of the broad \ha\ emission, which is not the case (see Sect.~\ref{s:blrs}). Consequently, we conclude that the presence of two BLRs is a more likely scenario.

Another option could be a single BLR affected by a blue-shifted \ha\ absorber, a scenario that has been observed in several high- and low-$z$ AGNs \citep[e.g.][]{Matthee24, Juodzbalis24, DEugenio25a, DEugenio25b, Ji25, Ji25b, Loiacono25, Lin25, Lin25b, Wang24, Ma25}. 
To investigate this, we considered a model including possible absorption of the continuum and broad emission, modelled with a Gaussian velocity distribution, assuming a standard parametrisation of 
\begin{equation*}
\small
    f_{\rm abs} = 1 - C_f + C_f \cdot \exp\left\{-\tau_0 \cdot \exp\left[-0.5\left(c\frac{1-\lambda_{\rm abs}/\lambda}{\sigma_{\rm abs}}\right)^2\right]\right\},
\end{equation*}
where $C_f$ is the covering factor of the absorber, $\tau_0$ is the optical depth at the line centre, $c$ is the speed of light, $\lambda_{\rm abs}$ is the centroid of the absorption, and $\sigma_{\rm abs}$ is the effective velocity dispersion of the absorbing gas \citep[see][]{Draine11, Juodzbalis24, DEugenio25a, DEugenio25b, Ji25, Loiacono25}.

To explore this scenario, due to the known parameter correlations in the absorption model \citep{DEugenio25b}, we adopt a Bayesian framework following the methods outlined by \citet{DEugenio25b}. 
We modelled simultaneously the region around \ha, as well as adjacent spectral windows (including \hb, \oiiill, \oiop, \niill, \siid). The narrow lines share the same redshift and velocity dispersion, which helps to constrain the redshift and dispersion of \ha\ in the presence of an absorber. 
We obtain a best fit with an absorber blue-shifted by 100~km/s, with $\sigma_{\rm abs}=88$$\pm$$4$~km/s, $C_f$$=$$0.996^{+0.003}_{-0.008}$, and $\tau_{\rm 0,H\alpha}$$=$$3.1^{+0.3}_{-0.2}$. For this model, the (single) central black hole has a mass of $\log(M_\bullet/M_\odot)$$=$$7.6$, accreting at 20 per cent of its Eddington limit. These parameters are comparable to other results for little red dots \citep[e.g.][]{DEugenio25a}. 
However, absorption from the unstable $n=2$ energy level requires extremely high column densities of neutral hydrogen and, effectively, extremely high densities, comparable to what is typically inferred for the BLR \citep[e.g.][]{Juodzbalis24}. For this reason, it is generally thought that the hydrogen absorber should be located within or immediately outside the BLR, i.e.~on sub-parsec scales. In this case, we can discard this model on the same grounds as the asymmetric single BLR model: we would expect no change of the centroid when sampling the BLR over different spectral windows.  However, for completeness, we calculate $\Delta\,\mathrm{BIC}$ for this model and the corresponding fiducial model of two central BLRs, and find that the latter is preferred with $\Delta\,\mathrm{BIC}\sim30$.

Could we see a single BLR in combination with a redshifted outflow in the centre? Outflows in distant galaxies are more typically observed symmetric around the ISM emission or blue-shifted due to differential extinction, but red-shifted emission associated with outflows has also been reported \cite[e.g.][]{Carniani24}. Interestingly, the majority of the low- to moderate-luminosity AGNs discovered with JWST at $z$$\sim$4-6 lack the strong outflow emission typically associated with more luminous quasars \citep[e.g.][]{Maiolino24}.
The broad component detected in \ha\ does not have a counterpart in the forbidden \oiii\ lines, which however feature a narrow component more than twice as bright as the narrow \ha\ component. We can therefore rule out an outflow origin of the secondary broad \ha\ component in the centre with metallicity and density similar to the galaxy ISM. However, the \oiii\ component of an outflow could be suppressed due to high densities. Assuming a gas phase metallicity of 0.1 solar or higher, we calculate a minimum density of $n_e>10^6$~cm$^{-3}$ to bring the non-detection of a broad \oiii\ component in line with the (secondary) broad \ha\ flux, based on calculations with \texttt{pyneb}. At this density, we estimate a gas mass of about 2000 solar masses from the (secondary) broad \ha\ flux, which should be unstable to gravitational collapse. Alternatively, the putative outflowing gas could have more typical ISM density, but would then require a metallicity significantly lower compared to the galaxy ISM. 
We therefore deem the outflow scenario highly unlikely.

\subsection{Supernovae, shocks, winds, and massive stars}\label{s:sn} 
Core-collapse supernovae (CCSNe) produce broad Balmer emission lines, although those are typically characterised by much broader line profiles \citep[several 1000 to 10000~km/s;][]{Taddia13, Gutierrez17}. These broad-line components are often blue-shifted or show P Cygni features, which is not the case for \tar, neither in the centre nor in the outskirts.
As discussed in detail by \cite{Maiolino25a}, average individual SNe may reach broad \ha\ luminosities of $10^{41}$~erg/s \citep[e.g.][]{Pastorello02, Taddia13} or up to $10^{42}$~erg/s in extreme cases. For the AGN in \tar, we measure broad \ha\ luminosities of $\log(\rm H\alpha_{\rm BLR}/({\rm erg/s}))=42.7\pm0.1$, 42.0$\pm$0.1 and 41.2$\pm$0.2, i.e.~above the typical luminosity for individual SNe, though in reachable range for the two less luminous broad components. 

For the two central AGNs we can leverage the existing NIRCam grism observations of \tar\ obtained on 14 January 2023 to compare the broad \ha\ luminosities over a time span of about 120 days rest-frame (23 month observed frame). If the central BLRs in \tar\ are indeed SNe initially observed at peak luminosity, over a course of about 120 days we may expect a drop in broad \ha\ luminosity of about a quarter to half an order of magnitude \citep[although individual cases can vary; see][]{Pastorello02, Taddia13}. 
We re-fit the lower-resolution, one-dimensional spectrum by \cite{Matthee24} with a two-BLR model and measure $\log(\rm H\alpha_{\rm BLR}/({\rm erg/s}))=42.6\pm0.1$ and $42.0\pm0.5$, i.e.~consistent with the values derived from our NIRSpec-IFS data. We therefore deem the SN scenario for the two central broad \ha\ components unlikely. 

Unfortunately, we cannot perform the same test for the NW broad \ha.
Instead, we explore more generally whether the broad \ha\ emission at the NW location could be due to shocks. Specifically, we consider the shock models by \cite{Allen08}, as implemented in MAPPING V \citep{Sutherland18}, with the following parameters: a pre-shock density of 1/cm$^3$, magnetic field strengths [0.0001-10]$\mu$G, and solar and SMC metallicity. Based on the line width of the broad \ha\ component, we assume a shock velocity of 1000~km/s \citep[see][]{Perna20}. The \ha\ luminosities predicted by these models (up to $\sim1.7\times10^{40}$~erg/s) are lower compared to the broad NW \ha\ component. Furthermore, these models predict emission line ratios incompatible with what we observe. As an example, the predicted [O~{\sc{i}}]$_{6300}$/\ha\ ratio is $\gtrsim0.3$. We do not detect [O~{\sc{i}}]$_{6300}$ in the NW spectrum extracted from the pink aperture, and also from the main galaxy aperture we find [O~{\sc{i}}]$_{6300}$/\ha$\ll0.3$ for any of the broad \ha\ components. 
We note that also other features typical of CCSN are not seen throughout \tar, such as a broad emission bump around 4600~\AA\ \citep{GalYam12}.
 
Localised starburst and very massive stars can also produce broad Balmer lines. Could these explain the broad \ha\ in the off-nuclear region? Star-formation-driven winds in high-$z$ galaxies have typical FWHM of $\sim400-500$~km/s \citep[see e.g.][]{FS19, Carniani24}, while we find a broad \ha\ FWHM of $>1000$~km/s. More importantly, if the broad \ha\ would be due to outflows (star-formation-driven or AGN-driven), we would expect to see corresponding broad emission in \oiiill\ \citep[e.g.][]{Harrison16, Concas22, Cresci23, Bertola25, Venturi25}, which is not observed for \tar.
In the case of very massive stars, the associated equivalent widths of \ha\ fall typically below 20~\AA\ after $\sim10$~Myr \citep{Martins20}.
For the \ha\ BLRs in \tar\ we find high equivalent widths of \ha\ ($\log(\rm EW_{H\alpha}/$\AA$)=2.7\pm0.3, 2.0\pm0.3$ and $2.5\pm1.2$). Therefore, we deem also this scenario unlikely.\\

To summarise, considering various physical scenarios, we conclude that we find evidence of three distinct Type~I AGNs in \tar: two at close separation in the centre, and one in the outskirts of \tar\ in the north-west. In both regions we observe broad \ha\ emission without a counterpart in the forbidden lines, and spectral features typically associated with shocks or supernovae are absent. Specifically in the centre, we find evidence of two compact sources based on their distinct line maps through spectro-astrometry, coinciding with an asymmetric integrated line shape consistent with a superposition of two BLRs.
For the reminder of this paper, we therefore proceed with the discussion of three massive, active black holes in \tar.

\section{Derivation of global galaxy properties}\label{s:global}

\subsection{SED fitting}\label{s:lmstar}

To constrain the global properties of \tar, we performed a spectral energy distribution (SED) fitting analysis using \texttt{CIGALE} \citep{Boquien19}. We leveraged the PRISM spectra to extract the photometry directly from the continuum emission. 
First, we masked any strong emission lines due to the uncertain contribution of the AGN. Then, we sampled the PRISM continuum in nine bins of 4000~\AA, which provides a good representation of the continuum shape without being too sensitive to noise fluctuations.

The SED fitting was performed considering a large parameter grid and taking into account the AGN emission. 
We used delayed star formation history (SFH) models which are able to reproduce both early-type and late-type galaxies, with an additional term that allows for a recent, constant burst in star formation. We adopted stellar templates from \cite{bruzual03}, and a \cite{Chabrier03} initial mass function. We considered two different metallicities for the stellar populations (solar and 0.2 solar) and a separation between the young and old stellar populations between 10 and 500~Myr. We also included the nebular emission module with gas phase  metallicities of solar and 0.2 solar. This module was computed by \texttt{CIGALE} in a self-consistent manner using a grid of \texttt{Cloudy} photoionisation calculations \citep{Ferland13}. For the attenuation of the stellar continuum emission, we used the \texttt{dustatt\_modified\_CF00} module \citep{Charlot00}, which allows different attenuations for the young and old stellar populations. We selected a narrow range of $A_{V,{\rm ISM}}$, the attenuation of the stellar continuum, to match the result obtained from the spectral analysis assuming $E(B$$-$$V)_{\text{star}}$$=$$0.44 E(B$$-$$V)_{\text{gas}}$. Assuming $E(B-V)_{\text{star}}$$=$$E(B$$-$$V)_{\text{gas}}$, as may be appropriate for young stellar populations \citep[see][]{Puglisi16, Perna25b}, lowers our best estimate of the stellar mass by 0.1~dex. We also included dust emission in the infrared following \cite{Dale14}. 

For the AGN contribution we employed the \texttt{skirtor2016} \citep{Stalevski16} module updated by \cite{Yang20}, which has demonstrated to be reliable in studying various aspects of AGNs \citep[e.g.][]{Mountrichas22, Lopez23, Yang23, Mazzolari25}. The SED produced by the AGN combines emission from the accretion disc, torus, and polar dust. The accretion disc, responsible for the UV-optical emission in the central region, is parametrised according to \cite{Schartmann05}. 
Photons from the accretion disc can be obscured and scattered by dust in the vicinity, within the torus and/or in the polar direction. For the torus, the \texttt{skirtor2016} module employs a clumpy two-phase model \citep{Stalevski16}, based on the 3D radiative-transfer code SKIRT \citep{Baes11, Camps15}. For the polar dust component, we used the Small Magellanic Cloud (SMC) extinction curve, recommended for AGN observations \citep[e.g.][]{Bongiorno12}. For the extinction amplitude, we chose a range up to $E(B-V) = 1.5$. The code assumes energy conservation for both the polar dust absorbing the AGN emission and for the ISM dust, the former contributing to the AGN mid-infrared emission, the latter re-emitting isotropically following the prescriptions of \cite{Dale14}.
Finally, we allowed for inclination values from 20 to 90 degrees and varied the AGN fraction (defined as the ratio between the AGN luminosity and the total galaxy luminosity between 0.13 and 5\micron) from 0.1 (weak AGN contamination) to 0.9 (dominant AGN emission). 

We obtain a stellar mass estimate of $\log(M_\star/M_\odot)=9.1\pm0.5$, based on a circular aperture extraction with a diameter of 11 pixels ($r=1.7$~kpc). 
This aperture includes the one of the two small blue sources closer to \tar\ seen in the NIRCam imaging. The position of AGN~III is just at the boundary of the aperture. In Sect.~\ref{s:scaling} we discuss a separate stellar mass estimate for the NW black hole region.

\subsection{Integrated velocity dispersion and dynamical mass}\label{s:sig}

We measured the integrated ionised gas velocity dispersion from our fit to the narrow \oiii\ line emission from the main galaxy aperture spectrum (see Fig.~\ref{f:overview}), accounting for the LSF, and find $\sigma_{\rm int,[OIII]}=63$~km/s, adopting a floor uncertainty of 5~km/s. Following \cite{Uebler23}, we used the empirical relation between integrated velocity dispersion of ionised gas and stars found by \cite{Bezanson18b} to estimate the integrated stellar velocity dispersion $\sigma_\star$ for our target. We note that this conversion was established for $z\sim1$ galaxies with stellar masses $\log(M_\star/M_\odot)\gtrsim9.9$; and yet, no systematic investigation of this type exists for higher-$z$ galaxies \citep[for AGN-specific studies in the local Universe see][]{Bennert18, Remigio25}. For \tar, we find $\sigma_\star\sim83$~km/s (i.e.~$\log(\sigma_\star/({\rm km/s}))\sim1.92$). We adopt an uncertainty of 0.10~dex to account for the scatter found in the relation by \cite{Bezanson18b}.

From our spaxel-by-spaxel fitting of the high-resolution data (see Sect.~\ref{s:blrs}), we find a fairly smooth velocity gradient of about $\Delta v\sim100$~km/s based on the narrow line emission across the main galaxy component (see Appendix~\ref{a:narrowmaps}). 
This may be used together with the velocity dispersion to constrain the galaxy kinematics and to construct a dynamical mass model. However, the possibility of an ongoing galaxy merger based on the NIRCam imaging and/or the evidence of a second massive black hole in the centre of \tar\ complicates the modelling. We defer a more detailed analysis of the galaxy kinematics of this source to future work. For the purpose of the present analysis, we attempt a simple first-order estimate of the total mass of \tar\ by approximating $M_{\rm dyn}$$\sim$$5\sigma_\star^2 R_e/G$ \citep{Cappellari06}. Here, $G$ is the gravitational constant, and $R_e$ is the effective radius which we estimate from the major axis of a two-dimensional Gaussian fit to the \oiii\ line map, correcting for the approximate PSF at this wavelength (PSF$_{\rm FWHM,[OIII]}$$\sim$$0.185^{\prime\prime}$) through subtraction in quadrature. With this approach, we find $R_e$$\sim$$0.64$~kpc
and $\log(M_{\rm dyn}/M_\odot)$$\sim$$9.7$. This estimate could be further refined through constraints on the structural parameters, specifically the S{\'e}rsic index $n$ and projected axis ratio $q$. Adopting common combinations of these parameters ($n$$\in$$[0.5; 8]$; $q$$\in$$[0.2;1]$) and using the calibrations by \cite{vdWel22}, our dynamical mass estimates for \tar\ vary in the range $\log(M_{\rm dyn}/M_\odot)$$\sim$9.5-10.2. In any case, our stellar mass estimate of \lmstar\ is lower than $M_{\rm dyn}$, allowing for 2.5 times or more additional mass in the form of gas and dark matter.

\section{The black holes in \tar}\label{s:bhprops}

\subsection{Masses}\label{s:mbh}

We used the single-epoch calibration by \cite{Reines13} to estimate the masses of the black holes in \tar\ based on the best-fit FWHMs and luminosities of the H$\alpha$ BLR components\footnote{\scriptsize 
Masses estimated from the single-epoch calibrations by \cite{Greene05} are about 0.2-0.4~dex (0.6~dex) lower using the H$\alpha$ (H$\beta$) BLR components for the AGN in \tar.} (without dust correction).
The corresponding measurements and results are reported in Table~\ref{t:bhs}. Listed uncertainties on black hole masses are dominated by the scatter in local scaling relations. We adopt a value of 0.4~dex \citep[e.g.][]{Ho14} which we add in quadrature to the propagated uncertainties from our best-fit results. 
We find black hole masses of $\log(M_\bullet/M_\odot)=7.9\pm0.4, 5.8\pm0.5$ and $6.3\pm0.5$ for the primary central AGN (I), the lower-mass central AGN (II), and the north-western AGN (III), respectively (or BH$_1$, BH$_2$, BH$_3$, respectively). If our interpretation is correct, BH$_2$ is among the lowest-mass black holes so far discovered above $z\sim5$ (see Sect.~\ref{s:scaling}). There is one AGN candidate by \cite{Maiolino24} with a lower mass estimate of $\log(M_\bullet/M_\odot)=5.65\pm0.31$ (they assume an intrinsic scatter of 0.3~dex for the local scaling relation), which was interestingly also identified as a possible secondary AGN in a system at $z\sim5.9$ through NIRSpec-MSA slit spectroscopy within the JADES programme. Our program BlackTHUNDER is carrying out NIRSpec-IFS observations of this target at both low and high spectral resolution, which will be crucial to confirm the nature of this system.

\begin{table*}
\small
\centering
\caption{Black hole properties.}
\begin{tabular}{lccccccc}
    & FWHM$_{\rm H\alpha,BLR}$  & $L_{\rm H\alpha,BLR}$  & $\log(M_\bullet/M_\odot)_{\rm H\alpha}$ & $\log(L_{\rm bol}/(\rm erg/s))_{\rm H\alpha,BLR}$ & $\log(\lambda_{\rm Edd})$ & $d_{{\rm I},i}$ & $\Delta v$ \\
    & [km/s] & [$10^{42}$ erg/s] & & & & [kpc] & [km/s] \\
\hline
    AGN I & $2920\pm140$ & $4.80^{+0.37}_{-0.38}$ & $7.9\pm0.4$ & $44.8\pm0.6$ & $-1.18\pm0.74$ & & $110$ \\
    AGN II &  $430\pm80$ & $1.03^{+0.22}_{-0.23}$ &$5.8\pm0.5$ & $44.1\pm0.6$ &$0.18\pm0.77$ & $0.19\pm0.04$ & $228$ \\
    AGN III & $1060\pm170$ & $0.16\pm0.04$ & $6.3\pm0.5$& $43.3\pm0.6$ & $-1.07\pm0.62$ & $\sim1.66$ & $3$ \\
\end{tabular} 
\tablefoot{
Properties of the massive black holes in \tar\ and related measurements from the high-resolution data. Measurements for the primary AGN~I and AGN~II are extracted from our best fit to the main galaxy aperture spectrum; properties for AGN~III are extracted from the best fit to the NW black hole aperture spectrum. The projected distances $d_{{\rm I},i}$ are with respect to the primary AGN~I. $d_{\rm I,II}$ is obtained from the channel map centroids (Sect.~\ref{s:blrs}), while $d_{\rm I,III}$ is an approximation. Relative velocities of the BLR components $\Delta v$ are reported with respect to the narrow \ha\ emission from the main galaxy aperture spectrum.
}
\label{t:bhs}
\end{table*}

While the stellar mass estimate for \tar\ is within the mass range of the local low-luminosity AGNs used to derive the scaling relation by \cite{Reines13}, it is not entirely clear whether such local scaling relations can be applied to higher-$z$ broad-line AGNs. There are indications that these estimates may carry systematic uncertainties of $0.5-1$~dex, as we detail in the following.
Recently, \cite{Greene25} argued that the bolometric luminosities of a sub-class of high-$z$ AGNs, the so-called little red dots \citep{Matthee24}, could be overestimated by an order of magnitude. 
Then, there are some indications that the geometry of many of the high-$z$ BLRs could be different: ubiquitous absorption features in the broad lines of hydrogen and helium as well as the lack of $X$-ray detections for the majority of these AGNs suggest that the covering factor of high-$z$ BLRs could be significantly higher than typical in local AGNs \citep[e.g.][]{Matthee24, Juodzbalis24, Ji25, Loiacono25, DEugenio25a, DEugenio25b, Maiolino25a}. This could introduce a systematic overestimation of the high-$z$ black holes masses by factors of two to three ($0.3-0.5$~dex in log).
Similarly, it has been pointed out that if the black holes are accreting at super-Eddington levels, their BLR sizes would be smaller, for a given black hole mass, leading to an overestimation of their masses when using standard single-epoch relations calibrated on sub-Eddington samples. This could reduce black hole masses by up to 0.5-1~dex and may at the same time explain their $X$-ray weakness \citep{Lupi24, Lambrides24, Inayoshi25a}. 
On the other hand, at (super-)Eddington accretion, radiation pressure can result in a lower effective gravitational force felt by the BLR clouds, and thus, a lower observed FWHM of the BLR lines, which in turn would lead to an underestimation of black holes masses \citep[see][]{Marconi08}.

To date, the most accurate black hole mass measurements at high redshifts are obtained by resolving the BLRs of luminous quasars through interferometry with VLTI/GRAVITY+. So far, the masses of four (and upper limit for one) $z$$\sim$$2$ black holes have been obtained with this method  (\citealp{Abuter24}; GRAVITY+ Collaboration, in prep.). These measurements indicate that the masses inferred from the single-epoch calibration by \cite{Reines13} applied to the \ha\ BLR emission could overestimate high-$z$ black hole masses by factors of two to ten. 
Interestingly, the authors also find that the quasars observed in their studies are accreting at highly super-Eddington rates, with a BLR size significantly smaller than expected from local relations. 
A recent study by \cite{Parlanti25b} of five $z$$\sim$$2$ AGNs suggests that the masses of black holes with lower accretion rates could still be estimated with local scaling relations.
\cite{Davies25} made the first measurement of a quasar accreting at super-Eddington rates
at $z$$\sim$$4$, finding a black hole mass one order of magnitude smaller compared to estimates from single-epoch scaling relations. This source has a strong outflow (resulting in a prominent, blue-shifted wing of H$\beta$), and if the scaling relations are applied only to the discy part of the BLR, they are in better agreement with the BLR dynamical measurement. 

A dynamical mass estimate (or lower limit) for the black hole in a low-luminosity AGN at $z$$=$$7.04$ has recently been obtained within the \bt\ programme, through a spatially resolved spectroscopic measurement of the lensed source Abell2744-QSO1 \citep{Juodzbalis25b}. In this case, the dynamical measurement is consistent with black hole mass estimates from single-epoch methods using the \ha\ or \hb\ BLR components \citep{Furtak24, Ji25, DEugenio25a, Maiolino25b}.

Recently, it has also been proposed that the broad-line shapes seen in compact, high-$z$ AGNs do not trace solely the BLR, but are broadened by various scattering mechanisms \citep{Rusakov25, Naidu25, Chang25}, such that the intrinsic, narrower BLR lines would correspond to much lower black hole masses, by up to two orders of magnitude. While this may be the case for some objects, \cite{Juodzbalis26} and \cite{Brazzini25} have shown that these proposed mechanisms do not generally apply to low-luminosity AGNs at high redshifts \citep[see also][]{Juodzbalis25b}. 

The focus of this work is not the accurate determination of the black hole masses, but the discovery of three massive, active black holes in a galaxy just 1.2~Gyr after the Big Bang.
Therefore, for the purpose of this paper, we proceed with the black hole masses derived following the local single-epoch method, but emphasise that these estimates come with substantial uncertainties.

\subsection{Luminosities and Eddington ratios}\label{s:lum}

We calculate the bolometric luminosities following Eq.~(6) by \cite{Stern12} which utilises the H$\alpha$ BLR luminosity, and find $\log(L_{\rm bol}/({\rm erg/s}))=44.8\pm0.6$, $44.1\pm0.6$ and $43.3\pm0.6$ for AGNs~I, II, and III. Our estimate for AGN~I is consistent with the value of $\log(L_{\rm bol}/({\rm erg/s}))=44.8$ derived by \cite{Matthee24} using another calibration based on the \ha\ BLR flux \citep{Greene05, Richards06}. We note that different calibrations utilizing the narrow line widths of  H$\beta$, for example, under the assumption that the narrow line emission is dominated by the AGN \citep[e.g.][]{Netzer19}, can yield bolometric luminosities of about one order of magnitude higher for AGN~I (see also~\citealp{Perna23, Uebler23}).

We infer the Eddington luminosity $L_{\rm Edd}$ of the black holes in \tar\ using our estimates of $M_\bullet$. We calculate Eddington ratios of $\log(\lambda_{\rm Edd})=\log(L_{\rm bol}/L_{\rm Edd})= -1.18\pm0.74, 0.18\pm0.77$ and $-1.07\pm0.62$ for AGNs~I, II, and III, respectively.
It is interesting to see that the lower-mass black hole in the centre shows super-Eddington accretion, while the larger primary black hole accretes at sub-Eddington levels. This is quantitatively similar to the results by \cite{Maiolino24} who also find that the secondary AGNs in their dual AGN candidate galaxies accrete at Eddington or super-Eddington rates, while the respective primary AGNs accrete at sub-Eddington rates.
From the theoretical point of view, some models studying the accretion rates onto massive black holes in binary systems find preferential accretion rates on the secondary (lower-mass) black holes \citep[e.g.][]{Farris14, Duffell20, Siwek23}; however others do not \citep[e.g.][]{Ochi05, Hanawa10, Bourne24}.
In the observational study by \cite{Maiolino24}, all high-$z$ AGNs with $\log(M_\bullet/M_\odot)\lesssim6$ accrete at or above the Eddington limit. As discussed by the authors, this may be due to selection effects, because it would be harder to detect smaller black holes accreting at low Eddington rates \citep[see also discussion by][]{Juodzbalis24b}. 
It is therefore worthwhile noting the sub-Eddington accretion rate of the off-nuclear $\log(M_\bullet/M_\odot)=6.2$ black hole in the north-west of \tar, opening up a regime of low black hole mass and low accretion rate, so far unpopulated by moderate-luminosity high-$z$ AGNs.

\subsection{Narrow-line ratio diagnostics}\label{nlr}

The classical narrow line ratio diagnostics involving the strong rest-frame optical lines \hb, \oiii, \ha, \nii\ and \siid\ routinely used to identify AGNs \citep{Baldwin81, Veilleux87} have been shown to no longer be effective for many high-$z$ AGNs, likely due to their low metallicities \citep[see][]{Kewley13, Kewley13b, Nakajima22, Kocevski23, Uebler23, Harikane23, Maiolino24, ArevaloGonzales25, Scholtz25}. Instead, AGNs start to overlap with star-forming galaxies, which also shift locus in these diagrams due to different ISM conditions and abundance compositions of ionizing massive stars \citep[e.g.][]{Shapley25}.
\tar\ falls into the regime of $z=0$ star-forming galaxies on the \oiii/\hb\ {\it vs.\ }\nii/\ha\ and \oiii/\hb\ {\it vs.\ }\siid/\ha\ diagnostic diagrams, and would therefore be misclassified as a star-forming galaxy if using these classical diagnostics. Only in the \oiii/\hb\ {\it vs.\ }\oiop/\ha\ diagram would \tar\ be classified as an AGN \citep[see also][]{Juodzbalis26}.

\cite{Mazzolari24} proposed utilising the \oiiia/\hg\ narrow line ratio together with other rest-frame optical line ratios to robustly identify AGNs even at low metallicities \citep[see also][]{Uebler24b}. 
Unfortunately in the case of \tar, \oiiia\ is not covered with the high-resolution grating, and blended with \hg\ in the prism observations.
 However, we can explore several rest-frame UV diagnostic diagrams (see Table~\ref{t:fluxes_main_prism}), but caution that in the case of \tar\ the uncertainties are large for all rest-frame UV diagnostics explored. In the \ciiil/\heiiuv\ {\it vs.\ }\civll/\ciiil\ diagram, \tar\ falls just at the separation of AGNs and composite galaxies based on the demarcation by \cite{Hirschmann19}. In both the \civ/\ciii\ {\it vs.\ }(\ciii+\civ)/\heiiuv\ and the EW(\heiiuv/\AA) {\it vs.\ } \ciii/\heiiuv\ or {\it vs.\ }\civ/\heiiuv\ diagrams, \tar\ falls into the regime populated by AGN models according to \cite{Nakajima22}.
These results underline the suitability of rest-frame UV line ratio diagnostics to identify AGNs also for high-$z$, low-metallicity black holes, although higher spectral resolution and higher $S/N$ observations would be helpful to explore the rest-frame UV emission of \tar\ in more detail.

\subsection{Scaling relations with host galaxy properties}\label{s:scaling}

\begin{figure}
\centering
        \includegraphics[width=0.95\columnwidth]{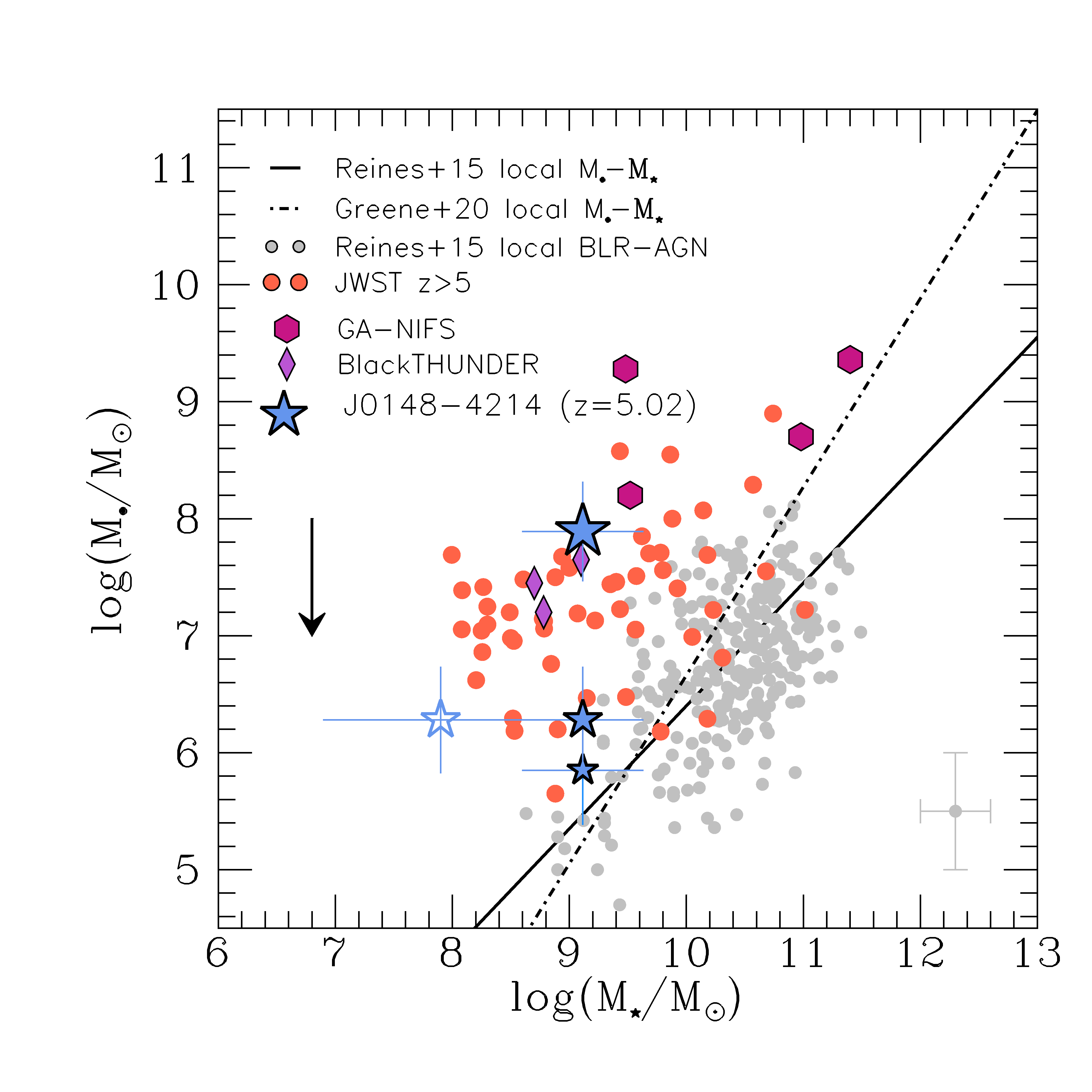}
    \caption{Black hole mass as a function of host galaxy stellar mass. We show the black holes in \tar\ as filled blue stars, other \bt\ sources as purple diamonds \citep{DEugenio25b, Maiolino25b, JonesG26}, sources from the GA-NIFS survey as pink hexagons \citep{Uebler23, Perna23, Parlanti24, Marshall25}, sources from the JADES survey \citep{Maiolino24, Juodzbalis26} and a number of other studies \citep{Tripodi24, Akins25, Maiolino23a, Rinaldi25, Taylor25, Zhang25} as orange circles, and data from local AGNs in grey \citep{Reines15}. Local relations by \cite{Reines15} and \cite{Greene20} are shown as solid and dash-dotted lines, respectively. The black arrow indicates a potential downward correction of one order of magnitude, based on considerations regarding the high-$z$ BLR geometry (see Sect.~\ref{s:mbh} for details). The primary black hole in \tar\ appears overmassive compared to the local relations, while the two lower-mass black holes in \tar\ fall within the scatter of the local relations. The open blue star indicates the position of the NW black hole if a separate estimate of the stellar mass of the north-western region of \tar\ is considered (see Sect.~\ref{s:scaling} for details).
    }
    \label{f:scaling_mstar}
\end{figure}

In Fig.~\ref{f:scaling_mstar} we plot the masses of the black holes in \tar\ as a function of stellar mass. Here, we initially use the total stellar mass derived from SED fitting as described in Sect.~\ref{s:lmstar} as a reference for all three black holes (filled blue stars). We compare our measurements to a sample of low-luminosity AGNs in the local Universe by \cite{Reines15} together with their scaling relation (grey points and black line), the local scaling relation by \citet[dash-dotted line]{Greene20}, and to a number of high-$z$ AGNs discovered or studied with JWST \citep{Uebler23, Perna23, Parlanti24, Tripodi24, Marshall25, Akins25, Juodzbalis26, DEugenio25b, Maiolino23a, Maiolino25b, Rinaldi25, Taylor25, Zhang25, JonesG26}.

As noted by many authors, the majority of JWST-discovered AGNs lie above the local scaling relations, although some sources fall within the scatter of local measurements \citep[e.g.][]{Harikane23, Kokorev23, Furtak24, Maiolino24, Juodzbalis26, LiR25}. The primary AGN of \tar\ (large filled blue star) has a similar offset from the local relation as other low-luminosity AGNs at similar redshifts. In contrast, the two lower-mass black holes in the centre and the NW black hole (smaller filled blue stars) fall within the scatter of local measurements when we use the total stellar mass of \tar\ as a reference.
However, the stellar mass associated with the NW black hole would be different if this is in the process of being accreted onto the main galaxy, for example through a minor merger. We attempt to constrain the stellar mass of the NW region specifically, by (i) repeating our SED fitting for an aperture extraction of three-by-three pixels centred on the NW black hole, and by (ii) assuming that the continuum tail seen in the NIRCam imaging is dominated by stellar continuum with the same mass-to-light ratio as the main galaxy, and scaling our total mass estimate based on the surface brightness ratio determined from the F115W imaging. While both of these approaches are strongly limited by the low $S/N$ in this region as well as PSF effects which are unaccounted for, we find approximate masses of $\log(M_\star/M_\odot)_{\rm NW}=8.0\pm0.6$ and $7.8\pm0.8$. The averages of these estimates and the corresponding position of the NW black hole are indicated in Fig.~\ref{f:scaling_mstar} as an open blue star. If the so approximated stellar mass does correspond to a companion galaxy, also the NW black hole should be considered overmassive with respect to local AGNs, although we caution that in this case some of the original stellar mass associated with the companion could have been tidally stripped \citep[e.g.][]{Yu02, Volonteri22}. 
On the other hand, if the NW black hole is instead in the process of being ejected from \tar\ following a three-body interaction with the two central black holes, or as a consequence of a previous massive black hole merger (i.e. a recoil black hole), assigning the main host stellar mass also to BH$_3$ is still reasonable (see also discussion in Sect.~\ref{s:fate} and the tentative kinematic evidence discussed in Appendix~\ref{a:narrowmaps}).
The secondary black hole in the centre may have formed in situ in \tar, in which case assigning the host stellar mass is reasonable. However, it could also have been accreted through a previous galaxy merger. Due to the small separation of the two central black holes there is no hope of identifying or determining a potential separate stellar mass component for BH$_2$ from current data. Considering also the smooth central ISM kinematics (see Appendix~\ref{a:narrowmaps}), we consider the secondary BH fully embedded in \tar\ at the time of observation. 
In the future, larger datasets of high-$z$ black hole multiplets could provide a way to statistically investigate their origins \citep[see e.g.][]{Lupi21, Zhang24}.

It has been pointed out that the prevalence of overmassive, moderate-luminosity black holes at high redshifts in the $M_\bullet-M_\star$ plane may be due to selection bias and sensitivity limits, coupled with a higher dispersion of the relation (\citealp{Li25}; see also \citealp{Silverman25} for luminous quasars). 
Stacking exercises for sources without individual AGN detections (either Type~I or Type~II) at $3<z<7$ have revealed average black hole masses much closer to local relations \citep{Geris25}. 
A recent study by \cite{Ren25} finds one robust and six candidate Type~I AGNs in $z\sim4.4-5.7$ main-sequence galaxies from the ALPINE-CRISTAL-JWST survey \citep{Faisst26} which are consistent with local relations. 
Certainly, there exists a large and likely dominant population of lower-mass black holes at high redshifts, which has not been detected in current JWST datasets due to selection biases and sensitivity limitations. Such a population has been modelled by cosmological simulations of galaxy formation \citep[e.g.][]{Sijacki07, Sijacki15, Volonteri20, Jeon23, Habouzit22, Habouzit25}.
However, even if the offset of the $M_\bullet-M_\star$ relation at high $z$ is dominated by selection effects, the finding of overmassive black holes (even when accounting for possible systematics; see Sect.~\ref{s:mbh}) which constitute a significant fraction of their entire host galaxy mass, like the primary AGN in \tar, is remarkable. This population provides informative constraints on possible black hole seeding and growth mechanisms in the early Universe \citep[e.g.][]{Schneider23, Natarajan24, Regan24}.
Finally, we note that overmassive black holes have been detected to exist also at later cosmic times \citep[e.g. around $z\sim0.5-3$;][]{Mezcua23, Mezcua24}.

\begin{figure}
\centering
    \includegraphics[width=0.95\columnwidth]{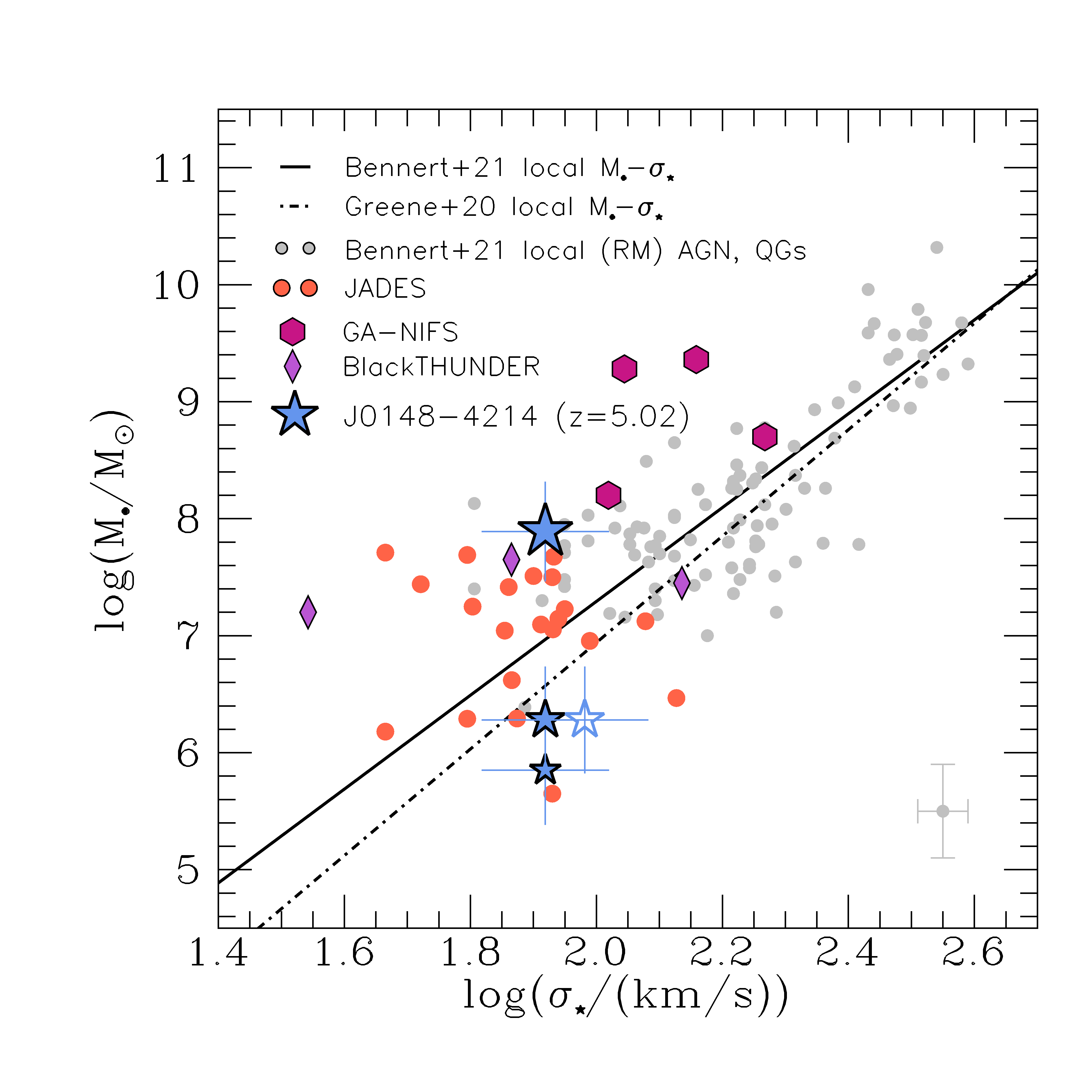}
    \caption{Black hole mass as a function of stellar velocity dispersion. We show the black holes in \tar\ as filled blue stars, other \bt\ sources as purple diamonds \citep{DEugenio25b, Maiolino25b, JonesG26}, sources from the GA-NIFS survey as pink hexagons \citep{Uebler23, Perna23, Parlanti24, Marshall25}, sources from the JADES survey as orange circles \citep{Maiolino24, Juodzbalis26}, and data from local AGNs and quiescent galaxies (QGs) in grey \citep{Kormendy13, Bennert21}. Local relations by \cite{Bennert21} and \cite{Greene20} are shown as solid and dash-dotted lines, respectively. For high-$z$ literature sources, we (re-)calculated $\sigma_\star$ according to \cite{Bezanson18b} as described in Sect.~\ref{s:sig}. The black holes in \tar\ scatter around the local relations, even when a separate $\sigma_\star$ estimate is considered for the NW black hole (open blue star; see Sect.~\ref{s:scaling} for details).} 
    \label{f:scaling_sigs}
\end{figure}

The black holes in this study exist just 1.15~Gyr after the Big Bang. We can expect that they and their host galaxy will grow by a substantial amount during the more than 12~Gyr time span until the present day. The comparison with local, moderate-luminosity AGNs therefore serves merely to highlight any differences of these early AGNs with local, similarly active galaxies. 
It is likely that our high-$z$ AGNs will grow into massive, evolved galaxies by $z=0$. 
In the local Universe, the black hole masses of such systems are typically compared to the stellar mass or velocity dispersion of the bulge component (with bulge-to-total fractions reaching up to $B/T=1$) of their host galaxies \citep[e.g.][]{Ferrarese00, Kormendy13, McConnell13, Bennert15}. 
In Fig.~\ref{f:scaling_sigs} we plot the masses of black holes in \tar\ as a function of stellar velocity dispersion (see Sect.~\ref{s:sig}). Again, we initially use the value for the main galaxy as reference for all black holes in \tar, but we also show a separate estimate for the NW black hole (open blue star), based on $\sigma_\star$ inferred from the \oiiil\ line from the NW extraction (see Sect.~\ref{s:sig} for the assumptions involved).\footnote{\scriptsize 
Using instead the \ha\ narrow line measured from the NW extraction ($\sigma_{\rm H\alpha,n}=14^{+16}_{-14}$~km/s) would result in a poorly constrained value of $\log(\sigma_\star/{\rm (km/s)})\sim1.35\pm0.51$, arguably too low compared to the spectral resolution of NIRSpec to be trustworthy.}
\cite{Maiolino24} pointed out that the majority of high-$z$ AGNs that appear overmassive in the $M_\bullet-M_\star$ plane are more consistent with local galaxies in the $M_\bullet-\sigma_\star$ plane \citep[see also][]{Juodzbalis26}. Also the black holes in \tar\ are spread around the low-$\sigma_\star$ end of the local relations by \cite{Bennert21} and \cite{Greene20}.
This could indicate that high-$z$ galaxies such as \tar\ evolve along the $M_\bullet-\sigma_\star$ relation until the present day \citep[see also][]{Jahnke11}. 
Since the integrated velocity dispersion is a proxy for the total dynamical mass of a system, it may also indicate that the total mass typically associated with local (quiescent) black holes is already in place at early times, in the form of gas and/or dark matter \citep[e.g.][]{McClymont25}.
This would imply that, on average, galaxies (not just their stellar content) and massive black holes do grow together over most of cosmic history \citep[see also][]{Trinca24}. Ideally, independent measurements of the galaxy gas mass would help to further investigate the relation between the total baryon content and black hole mass at these early cosmic epochs.

We also note that the increase in scatter seen at the lower end of the $M_\bullet-\sigma_\star$ relation could be a result of the highly dynamic evolution of massive black hole binaries or multiplets at early cosmic times, which may drive galaxies temporarily off the main relation. For instance, the ejection of a massive black hole would move a galaxy below the relation \citep[e.g.][]{Mannerkoski22}, while the potential capture of a massive, solitary black hole could move a galaxy above the relation.
The latter, however, is certainly a less likely event, and other physical processes or larger intrinsic uncertainties should be considered as possible reasons for the increased scatter above the relation at the low-$\sigma_\star$ end. For instance, a smaller scatter at the high-mass end could simply be the result of a higher number of black hole mergers \citep{Hirschmann10}. 
We note, however, that recently a population of ultra-massive black holes has been identified in the local Universe, which are positive outliers on the high-mass end of the $M_\bullet-\sigma_\star$ relation \citep{Saglia24, deNicola25}. Clearly, the topic of the scatter in the $M_\bullet-\sigma_\star$ relation requires further investigation.
Ultimately, statistical investigations of the scatter at the low-$M_\bullet$ end may provide insights into the formation channels of black holes.

\subsection{What is the fate of the massive black holes in \tar?}\label{s:fate}

Close-separation dual or triple black holes such as those discovered in this work are of great interest in the context of black hole growth, since they are likely going to merge by $z=0$.
From the theoretical perspective, massive black hole triplets and their subsequent mergers have been studied in a number of models and across a range of scales \citep[e.g.][]{Volonteri03, Hoffman07, Kulkarni12, Bonetti18, Ryu18, Mannerkoski21, Partmann24, Buttigieg25, Rantala25}.

We first consider the massive primary black hole together with the lower-mass black hole in the central region of \tar.
From mapping the emission of the two BLRs of these black holes, we derived a projected separation of $190\pm40$~pc (Sect.~\ref{s:blrs}).
We note that the sphere of influence of the primary black hole is about $r_h\sim50$~pc (with $r_h=GM_\bullet/\sigma_{\star,\rm bulge}^2$, assuming $\sigma_{\star,\rm bulge}=\sigma_\star$; see Sect.~\ref{s:sig});  this means that for a circular orbit the two central black holes are not (yet) gravitationally bound, although this may be different in the case of an eccentric orbit.
Even if unbound, at these separations and considering the total stellar mass and size of the system, the two black holes are likely already in the efficient dynamical friction phase \citep{Chandrasekhar43, Ostriker75, Begelman80, OstrikerE99}. This process, during which the interaction of the black holes with the stars, gas and dark matter around them promotes the loss of angular momentum, precedes the later stages of massive black hole mergers -- the stellar hardening phase, reached at a separation of about $a_h=GM_{\bullet,2}/(4\sigma_\star^2)\sim0.1$~pc for the two central black holes in \tar\
(\citealp{Begelman80, Quinlan96}; $M_{\bullet,2}$ being the mass of the lighter black hole), and eventually the emission of gravitational waves leading to the coalescence of the black holes \citep{Peters63, Peters64}.

We use our black hole mass estimates together with the total stellar mass of \tar\ extracted from a 1.7~kpc radius aperture ($\log(M_\star/M_\odot)$$=$9.1; Sect.~\ref{s:lmstar}), and assume an absolute separation of the two central black holes of $0.3$~kpc. 
Following \cite{BT08}, we calculate the dynamical friction time for the secondary black hole BH$_2$ with mass $M$ to reach the primary one as $t_{\rm df}$$=$$1.65/\ln(\Lambda)\cdot (r_i^2\sigma_\star)/(G M)$, with $r_i$$=$0.3~kpc, and $\Lambda$$\approx$$b_{\rm max} v_{\rm typ}^2/(G M)$. Here, $v_{\rm typ}^2$$\approx$$G M_\star/R_e$, and we assume $b_{\rm max}$$=$0.3~kpc. 
We find an approximate inspiral time of $t_{\rm df}$$\sim$0.66~Gyr (corresponding to $z$$\sim$3.5).
We emphasise that this simplistic calculation neglects contributions from gas and dark matter which may shorten the dynamical friction time. If we use $M_{\rm dyn}$ instead of $M_\star$ in the above calculation, we find $t_{\rm df}$$\sim$0.55~Gyr ($z$$\sim$3.65); however, we stress that neither approach properly accounts for the effects of the dissipative gas component.
The off-nuclear black hole may also be in the process of sinking towards the galaxy centre (we discuss alternative scenarios further below). Using the same approach as above, and considering $r_i$$\sim$$b_{\rm max}$$\sim$1.7~kpc and our estimate for $M_\star$, we find an approximate inspiral time of $t_{\rm df}$$\sim$5.9~Gyr (corresponding to $z$$\sim$0.75). If we use $M_{\rm dyn}$ instead, we find $t_{\rm df}$$\sim$5.0~Gyr ($z$$\sim$0.93).
We note that the detection of not just two but three massive black holes in such close proximity is relevant in this context, since triple black hole interactions provide a mechanism to promote and accelerate the merging of massive black holes \citep[e.g.][]{Saslaw74, Hills80, Blaes02, Colpi02, Ryu18, Bonetti19, Mannerkoski22, Sayeb24}.

Assuming that the black holes will merge by the present day after the dynamical friction phase, the eventual massive black hole mergers in descendants of systems such as \tar\ may be detectable by the next-generation gravitational wave detector Laser Interferometer Space Antenna \citep[LISA;][]{Thorne76, AmaroSeoane23}. 
At the time of observation, the mass ratios $q=m_i/m_j$ for the black holes in \tar\ are $q=0.008$ for BH$_1$ and BH$_2$ and $q=0.025$ for BH$_1$ and BH$_3$ (or BH$_{1+2}$ and BH$_3$). The corresponding chirp masses $\mathcal{M}=(m_i\cdot m_j)^{3/5}/(m_i+m_j)^{1/5}$ (assuming no significant mass growth prior to coalescence) are $\mathcal{M}=10^{6.6}$ for BH$_1$ and BH$_2$ and $\mathcal{M}=10^{6.9}$ for BH$_1$ and BH$_3$ (or BH$_{1+2}$ and BH$_3$).
To obtain some first-order constraints on the potential detection probability with LISA, we use the simulator \texttt{GWFish} \citep{GWFish}. We emphasise that we make the simplifying assumptions of no further mass growth of the black holes, and that the black holes actually merge after the dynamical friction times calculated above. We further neglect spinning and precession of the black holes and apply the \texttt{IMRPhenomXHM} waveform model \citep{GarciaQuiros20}. As an additional caveat, we note that we treat the mergers of BH$_1$ with BH$_2$ and of BH$_{1+2}$ with BH$_3$ separately. In reality, as mentioned above, depending on the black hole separations and the masses of the black holes and host galaxy involved, it is also possible that the third black hole influences the gravitational wave signal of the binary \citep[e.g.][]{Bonetti19, Ficarra23}.
Sampling 100 realisations of our measured black hole masses within the quoted uncertainties, and random angular parameters, we obtain a detection probability of $38.2\pm4.2$ per cent for the merger of BH$_1$ and BH$_2$, and $81.8\pm1.8$ for BH$_{1+2}$ and BH$_3$, with a mean $S/N$ of $24.6\pm2.4$ and $279.5\pm11.8$, respectively. 
If we consider the shorter dynamical friction times obtained from using $M_{\rm dyn}$ instead of $M_\star$, we find somewhat reduced detection probabilities of $24.2\pm1.9$ and $78.2\pm0.8$ per cent.
While a number of simplifications and uncertainties are attached to these calculations, this highlights that the descendants of systems such as \tar\ may be plausible candidates for LISA detections.
In principle, the potential late-time mergers of massive black holes such as in \tar\ could also contribute to the gravitational wave background detected by the Pulsar Timing Array (PTA) collaborations \citep{NANOGravSMBHBs23, PTA2024, ETPA24}.

As mentioned earlier, an alternative fate for BH$_3$ may be the expulsion from the \tar\ system, considering that it could have been ejected from the centre through a three-body interaction with the other two massive black holes, or through a recoil kick as the consequence of a previous massive black hole merger \citep[e.g.][]{Peres62, Bekenstein73, Saslaw74, Mikkola90, Volonteri03, Madau04, Merritt04, Blecha16, Sijacki11, Mannerkoski22, DongPaez25}. In this scenario, if BH$_3$ were able to leave the parent halo of \tar, dragging some small amount of gas, stars, and dark matter with it, it could appear as a heavily overmassive, nearly solitary black hole at later times. This could resemble observations of high-$z$ massive black holes with little evidence of (much of) a host galaxy around them \citep[see][]{Chen25, Maiolino25b, Juodzbalis25b}. According to theoretical models, such solitary black holes resulting from three-body interactions may not be uncommon \citep{Volonteri03}. However, it is unclear whether the ejected black hole would be able to retain any additional material \citep[e.g.][]{Partmann24} and, therefore, specifically considering gas, whether it would be visible in accretion or not. If instead ejected through a gravitational wave recoil from a previous massive black hole merger, the retention of a compact, bound star cluster would be expected \citep[e.g.][]{OLeary09, Merritt04, Rawlings25}.
Since BH$_3$ is accreting material at the time of observation, a recoil event could therefore be more likely than a three-body interaction. 
The fact that BH$_3$ is not the lightest of the three black holes would further favour a recoil event over a three-body interaction \cite{Saslaw74}.
In this scenario, the $\gtrsim1.7$~kpc separation from the galaxy centre suggests at least a moderate kick velocity of $\gtrsim200$~km/s in order to exceed the galaxy escape velocity. This in turn implies a mass ratio of the progenitor black hole binary $q\gtrsim0.1$ \citep[e.g.][]{Tanaka09}. With a remnant mass of $\log(M_\bullet/M_\odot)\sim6.3$, this would therefore be an ideal LISA target, detectable at high redshifts with high $S/N$ \citep[see e.g.][]{Valiante21,Liu25}.

It is also a possibility that the NW black hole has been ejected previously, but will be re-accreted into the centre to eventually merge with the other two central black holes. From the theoretical perspective, such an evolutionary pathway has been described in detail in a cosmological context by \cite{Mannerkoski21}, predicting situations which are comparable to our observations in terms of the distances of the three black holes, although for black holes with higher and more similar masses ($\log(M_\bullet/M_\odot)\sim8.0-8.9$), and for later cosmic times ($z<1$).

\section{Conclusions}\label{s:conclusions}

In this work we have presented new NIRSpec-IFU data of the $z\sim5$ galaxy J0148-4214, revealing multiple \ha\ broad-line regions. Through detailed analysis of the spatially resolved data, and after discarding alternative interpretations such as supernovae, shocks, winds, or massive stars, we conclude that the observations are compatible with 
three massive ($M_\bullet\sim10^{5.8-7.9}M_\odot$) accreting black holes.
This discovery of a black hole triplet at high redshifts might indicate that multiple massive black hole systems are common in the early Universe.
In \tar, two of the black holes are already in close proximity, at a projected separation of about $190\pm40$~pc. While the integrated spectrum from the NIRCam grism observations showed an indication of a complex BLR profile, only the spectral mapping capabilities of the NIRSpec-IFU allowed us to disentangle the BLR profile and to separate the two black holes spatially. Similarly, the discovery of the third  faint black hole at a projected distance of about 1.7~kpc from the central black hole binary would not have been possible without the IFU. 
Additional data at higher spatial and spectral resolution will help to investigate in greater detail the dynamics of this system, and to achieve a refined characterisation of the central region of \tar.
Several integrated spectra of high-$z$ AGNs show BLR shapes comparable to \tar, indicating that they too could host dual or multiple AGNs. These systems require NIRSpec-IFU observations to be confirmed.
Our results highlight the crucial role of IFU observations in discovering massive black hole multiplets in galaxies in the early Universe, the progenitors of massive black hole mergers. Such black hole mergers, with the masses and timescales estimated in this work, may be detectable by next-generation gravitational wave observatories such as LISA.

\begin{acknowledgements}
We thank the anonymous referee for a constructive
report that helped to improve the quality of this manuscript.
We thank Stephanie Buttigieg, Filippo Mannucci, Thorsten Naab and Michael Kramer for valuable comments. 
We are grateful to Vardha Bennert for sharing a machine-readable version of Table~3 by \cite{Bennert21}.
H\"U thanks the Max Planck Society for support through the Lise Meitner Excellence Program.
H\"U and GM acknowledge funding by the European Union (ERC APEX, 101164796).
NMFS and GT acknowledge funding by by the European Union (ERC, GALPHYS, 101055023).
RM, FDE, LRI, and GCJ acknowledge support by the Science and Technology Facilities Council (STFC), by the ERC through Advanced Grant 695671 ``QUENCH'', and by the UKRI Frontier Research grant RISEandFALL. RM also acknowledges funding from a research professorship from the Royal Society.
SA acknowledges grant PID2021-127718NB-I00 funded by the Spanish Ministry of Science and Innovation/State Agency of Research (MICIN/AEI/ 10.13039/501100011033).
GC and EB acknowledge financial support from INAF under the Large Grant 2022 ``The metal circle: a new sharp view of the baryon cycle up to Cosmic Dawn with the latest generation IFU facilities'' and from the INAF GO grant 2024 ``A JWST/MIRI MIRACLE: MidIR Activity of Circumnuclear Line Emission''. EB also acknowledges support from the ``Ricerca Fondamentale 2024'' program (mini-grant 1.05.24.07.01). 
AJB acknowledges funding from the ``FirstGalaxies'' Advanced Grant from the European Research Council (ERC) under the European Union's Horizon 2020 research and innovation program (Grant agreement No. 789056).
SC and GV acknowledge support by European Union's HE ERC Starting Grant No. 101040227 - WINGS.
KI acknowledges support from the National Natural Science Foundation of China (12573015, W2532003), the Beijing Natural Science Foundation (IS25003), and the China Manned Space Program (CMS-CSST-2025-A09).
BL acknowledges the funding of the Deutsche Forschungsgemeinschaft (DFG, German Research Foundation) under Germany's Excellence Strategy EXC 2181/1 - 390900948 (the Heidelberg STRUCTURES Excellence Cluster).
MP acknowledges support through the grants PID2021-127718NB-I00, PID2024-159902NA-I00, and RYC2023-044853-I, funded by the Spain Ministry of Science and Innovation/State Agency of Research MCIN/AEI/10.13039/501100011033 and El Fondo Social Europeo Plus FSE+.
BRP acknowledges support from grants PID2021-127718NB-I00 and PID2024-158856NA-I00 funded by Spanish Ministerio de Ciencia e Innovación MCIN/AEI/10.13039/501100011033 and by ``ERDF A way of making Europe''.
RS acknowledges support from the PRIN 2022 MUR project 2022CB3PJ3 ``First Light And Galaxy aSsembly (FLAGS) funded by the European Union'' Next Generation EU, and from EU-Recovery Fund PNRR - National Centre for HPC, Big Data and Quantum Computing.
DS acknowledges support from the STFC, grant code ST/W000997/1.
RV and AT acknowledge support from PRIN MUR ``2022935STW'' funded by European Union-Next Generation EU, Missione 4 Componente 2 CUP C53D23000950006 and from the Bando Ricerca Fondamentale INAF 2023, Theory Grant ``Theoretical models for Black Holes Archaeology''.
SZ acknowledges the Texas Advanced Computing Center (TACC) for providing HPC resources under allocation AST23026.
Views and opinions expressed are those of the authors only and do not necessarily reflect those of the European Union or the European Research Council Executive Agency. Neither the European Union nor the granting authority can be held responsible for them.
This work is based on observations made with the NASA/ESA/CSA \textit{James Webb} Space Telescope. The data were obtained from the Mikulski Archive for Space Telescopes at the Space Telescope Science Institute, which is operated by the Association of Universities for Research in Astronomy, Inc., under NASA contract NAS 5-03127 for JWST. These observations are associated with programs \#1243 and \#5015 (DOI:10.17909/5rh4-pa55).
\end{acknowledgements}

\bibliographystyle{aa}
\bibliography{literature}


\begin{appendix}

\section{Full $R\sim2700$ spectrum of the main galaxy}\label{a:fullspec}

\begin{figure*}
    \centering  
    \includegraphics[width=\textwidth]{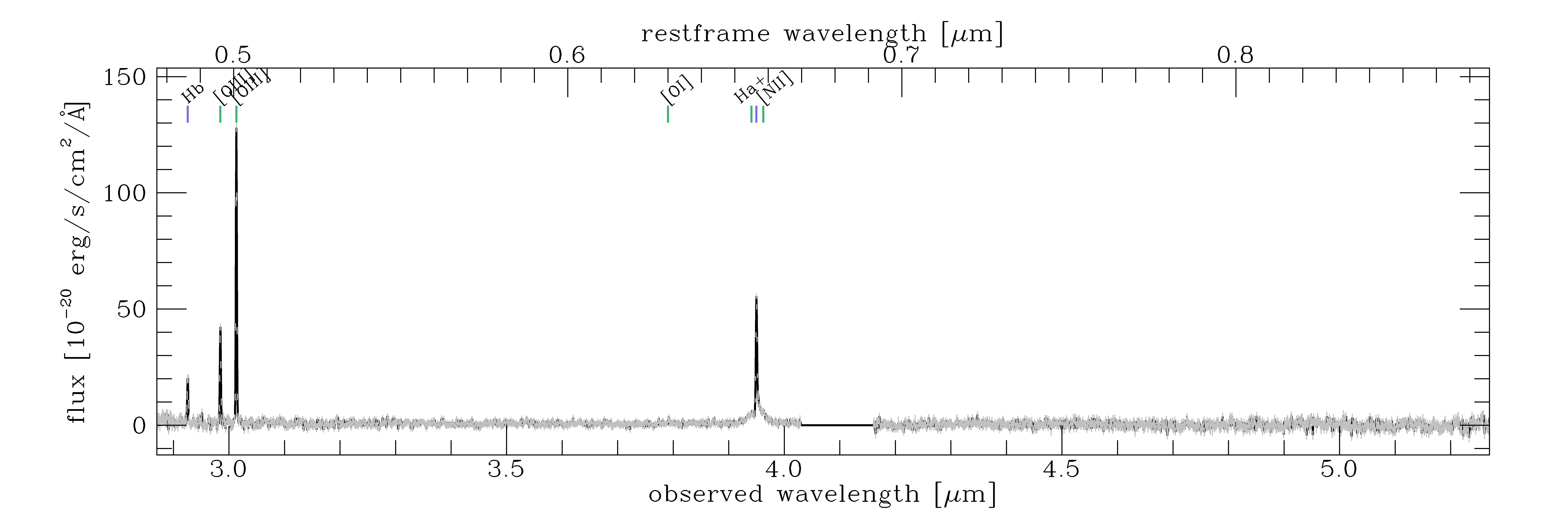}     
    \caption{Full $R\sim2700$ spectrum of the main galaxy (orange aperture in Fig.~\ref{f:overview}). We indicate the locations of several emission lines. The detector gap is masked in the region $\lambda_{\rm gap}\sim4.03-4.16\mu$m, where fewer than 15 slices contribute to the emission.}
    \label{f:fullspec}
\end{figure*}

In Fig.~\ref{f:fullspec} we show the full high-resolution spectrum integrated within the main galaxy aperture (without continuum subtraction). The locations of several emission lines discussed in Sect.~\ref{s:analysis} are indicated; we mask the detector gap where fewer than 15 slices contribute to the emission ($\lambda_{\rm gap}\sim4.03-4.16\mu$m). The continuum is not or only marginally detected in the high-resolution data.

\section{Channel maps of BLR1 and BLR2}\label{a:channelmaps}

\begin{figure}
    \centering  
    \includegraphics[width=0.49\columnwidth]{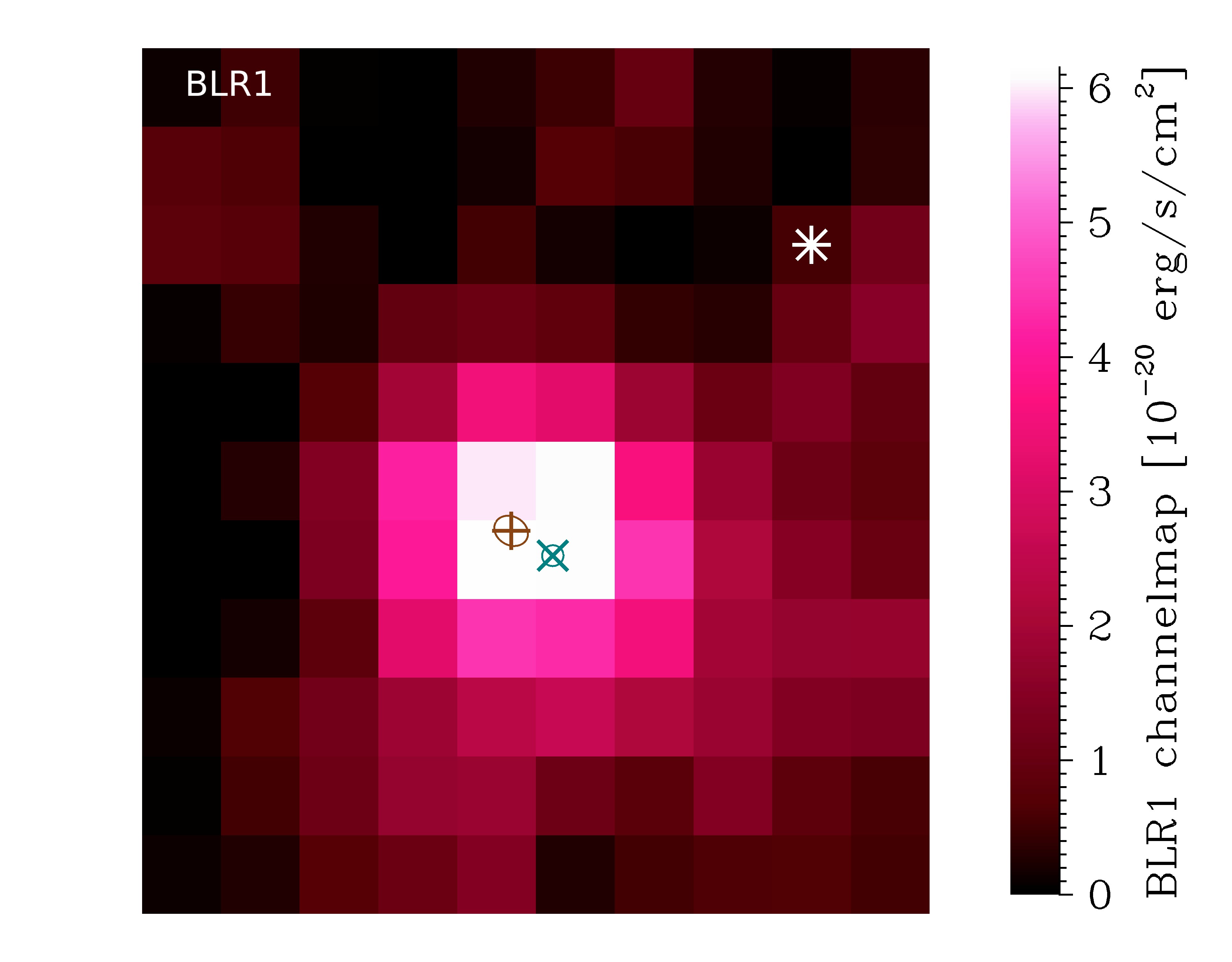}      
    \includegraphics[width=0.49\columnwidth]{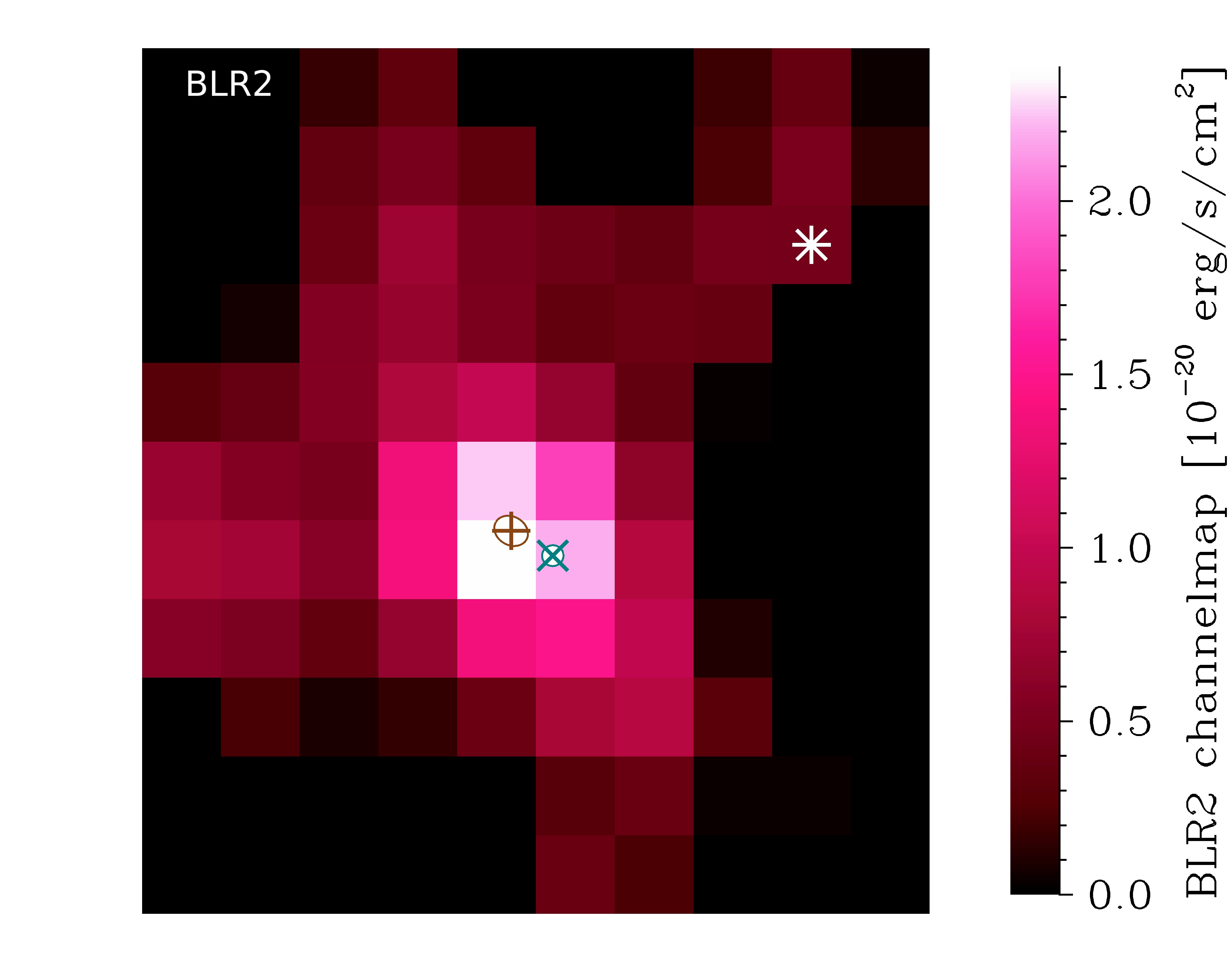}
    \caption{Channel maps of the two central broad \ha\ components, for BLR1 (\textit{left}) and BLR2 (\textit{right}), derived as described in Sect.~\ref{s:blrs}. The centroids of the channel maps obtained from two-dimensional Gaussian fits are indicated as a teal cross and a brown plus for BLR1 and BLR2, respectively; the   ellipses show two times the centroid uncertainty. The white star marks the position of the third BLR detected in \tar. The flux distributions of BLR1 and BLR2 are distinct, indicating that they do not originate from the same point source.}
    \label{f:channelmaps}
\end{figure}

In Fig.~\ref{f:channelmaps} we show the channel maps obtained for BLR1 and BLR2 as describe in Sect.~\ref{s:blrs}. The flux distributions of BLR1 and BLR2 are distinct, indicating that they do not originate from the same point source.

As described in Sect.~\ref{s:analysis}, we determine the centroids of BLR1 and BLR2 through two-dimensional Gaussian fits to the flux distribution of the two maps. Uncertainties are derived by varying the input images 10000 times each with their associated noise, repeating the fits, and adopting half the 68th percentile of the distribution as the uncertainties. 
The corresponding positions together with twice their fit uncertainty are indicated as a teal cross (BLR1) and a brown plus (BLR2) in Fig.~\ref{f:channelmaps}. We find positional uncertainties of 0.07 and 0.09 pixels. 
In general, the centroid of a point source can be determined with a precision of $\delta_{x,y}\sim0.4\cdot{\rm FWHM}/(S/N)$, where $S/N$ is measured at the peak \citep[e.g.][]{Neuschaefer95}.
In our case, we find $S/N\sim17$ and $S/N\sim11$ for the brightest $0.05\arcsec\times0.05\arcsec$ pixel of BLR1 and BLR2, indicating $\delta_{x,y}\sim0.09$ and $\sim0.14$ pixels, respectively. However, due to sampling several pixels contribute to the peak and the true $S/N$ and precision is expected to be better.
From the centroid measurements, we calculate the projected distance of BLR1 and BLR2 and find $d_{\rm I,II}=190\pm40$. In Appendix~\ref{a:blrtests} we demonstrate through mock tests that this distance and associated uncertainty are realistic, although the uncertainty may be underestimated by $10-20$~pc. We further note that, assuming the theoretically derived (upper limit) uncertainties on the positional measurements, we find a distance uncertainty of 50 pc.

The FWHM we determine from the Gaussian fits to the images are compatible with the expected FWHM of a point source observed with NIRSpec at the wavelength of \ha\ (ca.~0.197\arcsec, i.e.~3.94 pixel of 0.05\arcsec): we find $4.24\pm0.21$ and $3.64\pm0.73$ pixels for the average FWHM of BLR1 and BLR2 from our fits, where again the uncertainties are derived from our Monte Carlo runs.

\section{Mock tests concerning the separation of BLR1 and BLR2}\label{a:blrtests}

In Sect.~\ref{s:blrs} we describe how we measure the projected separation of BLR1 and BLR2 from channel maps, finding $d_{\rm I,II}=190\pm40$. To test whether such a positional uncertainty is reliable, we create two sets of 10000 two-dimensional Gaussians with a FWHM$=0.197$\arcsec (the PSF FWHM at the wavelength of \ha). This is to simulate the expected spread of a point-source (namely a BLR at cosmological distance) at the wavelength of \ha. The two sets of Gaussians are centred on the BLR1 and BLR2 centroids we measured from the channel maps in Sect.~\ref{s:blrs}, respectively. We add random noise, corresponding to the levels we measure in the channel maps of BLR1 and BLR2, respectively. We then measure the centroids of the resulting maps from a two-dimensional Gaussian fit and derive their distances. We find a median distance of $200^{+60}_{-40}$~pc, where the uncertainties correspond to the $68^{\rm th}$ confidence interval of the distribution. This suggests that the uncertainty of 40~pc on the projected distance of BLR1 and BLR2 we derived in Sect.~\ref{s:blrs} is plausible, although may be underestimated by $10-20$~pc.

\begin{figure}
\centering
        \includegraphics[width=0.7\columnwidth]{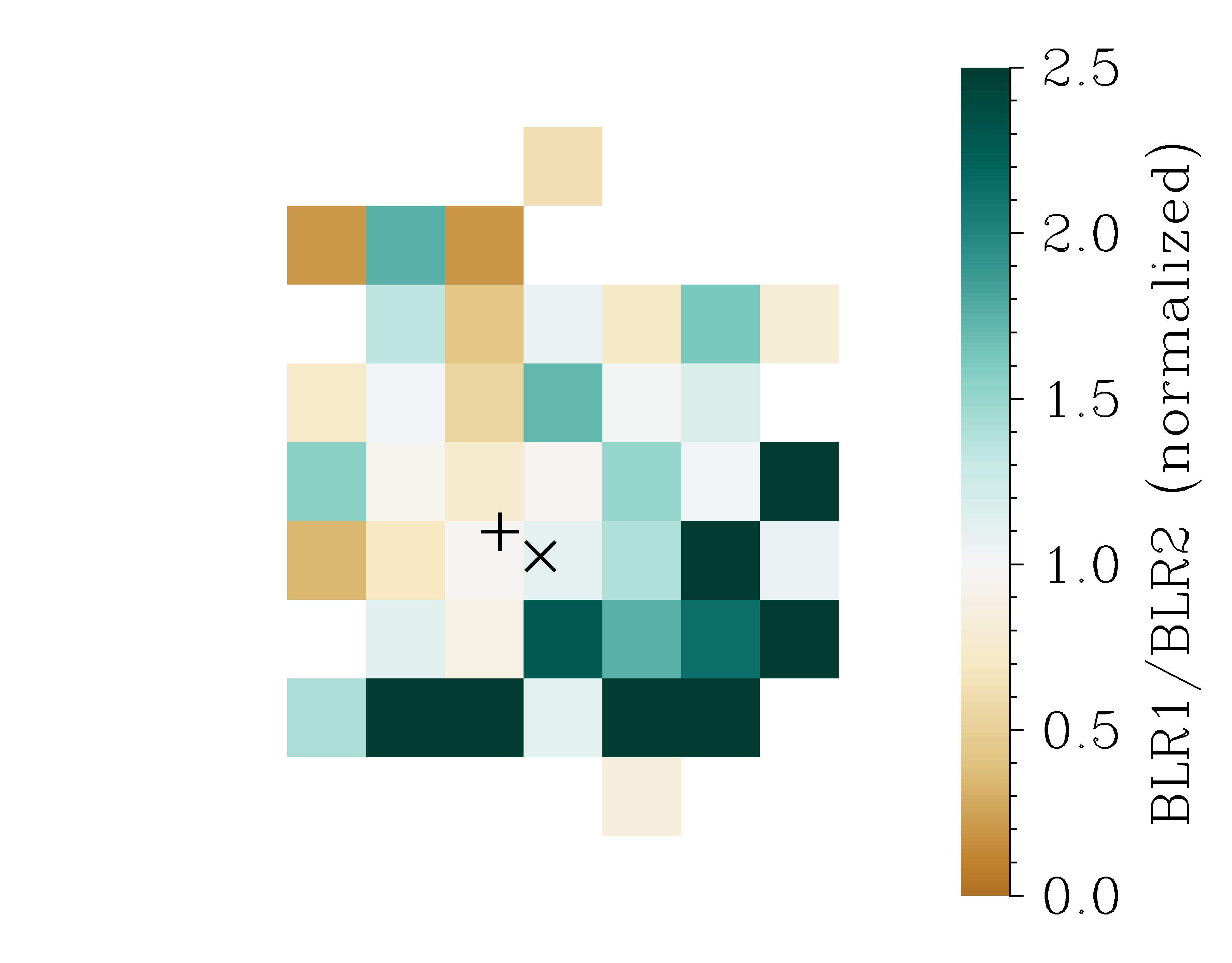}
    \caption{Ratio after peak flux normalisation of two BLRs from our mock sample. We plot pixels with $S/N\geq0.5$. A NE to SW modulation is visible, as expected for the images of two point sources spatially separated along that axis.}
    \label{f:blrratiotest}
\end{figure}

In Fig.~\ref{f:blrratiotest} we show the ratio after peak flux normalisation of two of these mock BLRs (i.e.~corresponding to the right panel in Fig.~\ref{f:blrmaps}). Like in the main part of the paper, we plot pixels with $S/N\geq0.5$. The NE to SW pattern obtained is qualitatively similar to what we found from our data in Sect.~\ref{s:blrs}. This further supports our interpretation that the two broad \ha\ components in the centre of \tar\ are indeed two point sources, separated by less than the PSF FWHM.

\section{$R\sim100$ fitting}\label{a:prism}

The prism observations covering the wavelength range 0.6\micron\ - 5.3\micron\ reveal many additional emission lines in the regime not covered by our high-resolution data.
We model narrow emission lines as Gaussians and tie the line centroids and narrow line widths for three groups of lines: (i) blue-wards of \oiill, (ii) from [O{\sc{ii}}] to \oiiill, (iii) from He~{\sc{i}}$_{\lambda5876}$ and red-wards. We also include one broad component for H$\alpha$ and for H$\beta$ (the resolution is too low to separate the two central BLRs spectrally). The continuum is modelled as a polynomial.

\begin{figure*}
        \includegraphics[width=\textwidth]{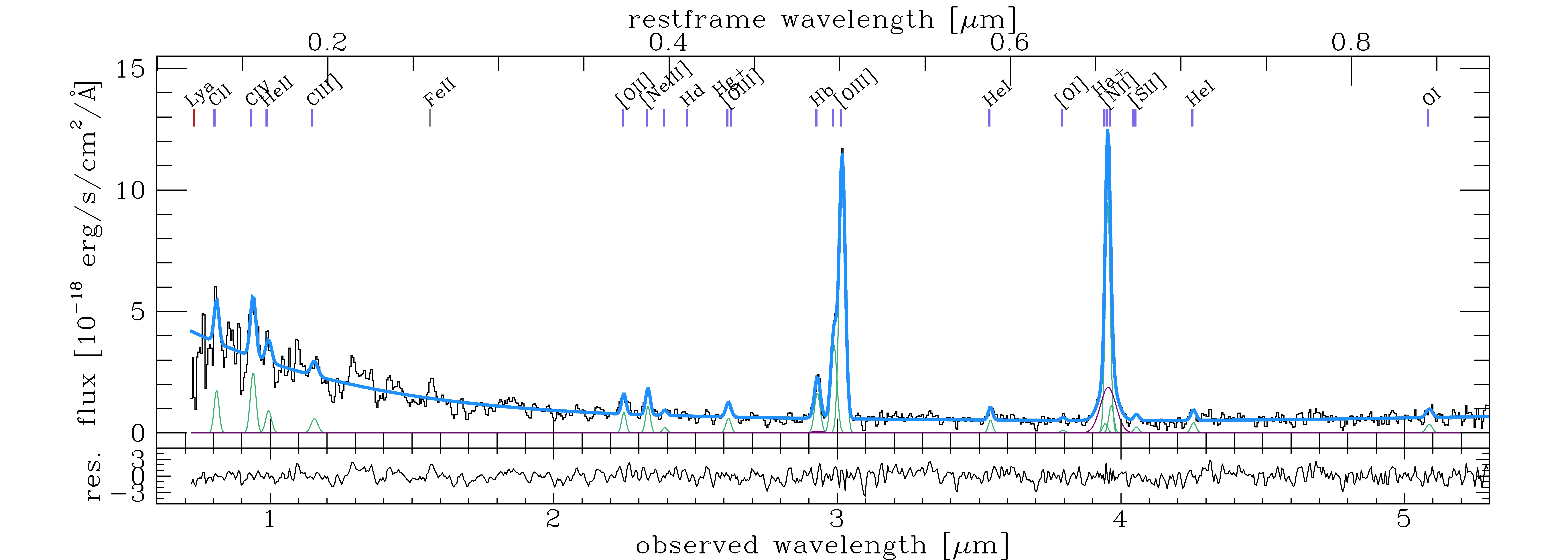}
    \caption{Fit to the prism spectrum extracted from the main galaxy (orange aperture in Fig.~\ref{f:overview}). The best fit is shown in blue, while the narrow and BLR components are shown in green and purple, respectively. Due to the lower resolution of the prism, the two-component structure of the \ha\ BLR seen in the high-resolution data cannot be resolved. We indicate the prominent emission lines. Ly$\alpha$ is not significantly detected. We indicate Fe~{\sc{ii}}$_{2599}$ coinciding with possible emission excess at the corresponding wavelength.}
    \label{f:prismfit}
\end{figure*}

We show our fit to the main galaxy aperture spectrum in Fig.~\ref{f:prismfit} and list the best-fit emission line fluxes in Table~\ref{t:fluxes_main_prism}. 
We note the detection of \oi\ in our prism aperture spectrum. This line has been detected in low-$z$ Seyfert~I, low-metallicity AGNs and quasars \citep[e.g.][]{Shields72, Netzer76, Grandi80, RodriguezArdila02, Matsuoka08, Izotov10} as well as in galactic emission nebulae \citep[e.g.][]{Morgan71, Muench74, Grandi75}. 
At high redshifts, it has been detected in a stack of $z\sim2-3$ star-forming galaxies from the CECILIA survey \citep{Strom23}, in the Sunburst Arc \citep{Choe25}, in a Wolf-Rayet galaxy from the MARTA survey \citep{Curti25}, and in a few $z\sim2-5$ AGNs \citep[][]{Juodzbalis24, Labbe24, DEugenio25b, Tripodi25}. 
The line-strength of \oi\ relative to \ha\ found for \tar, \oi/\ha$_{\rm narrow}=0.048$, \oi/\ha$_{\rm BLR}=0.076$ and \oi/\ha$_{\rm total}=0.029$ is comparable to those of local Seyfert~I galaxies (assuming negligible contamination from Paschen 17 and 18), while for the CECILIA star-forming sample and the galaxy from the MARTA program a sub-percentage ratio is found.

\begin{figure*}
\centering
        \includegraphics[width=\textwidth]{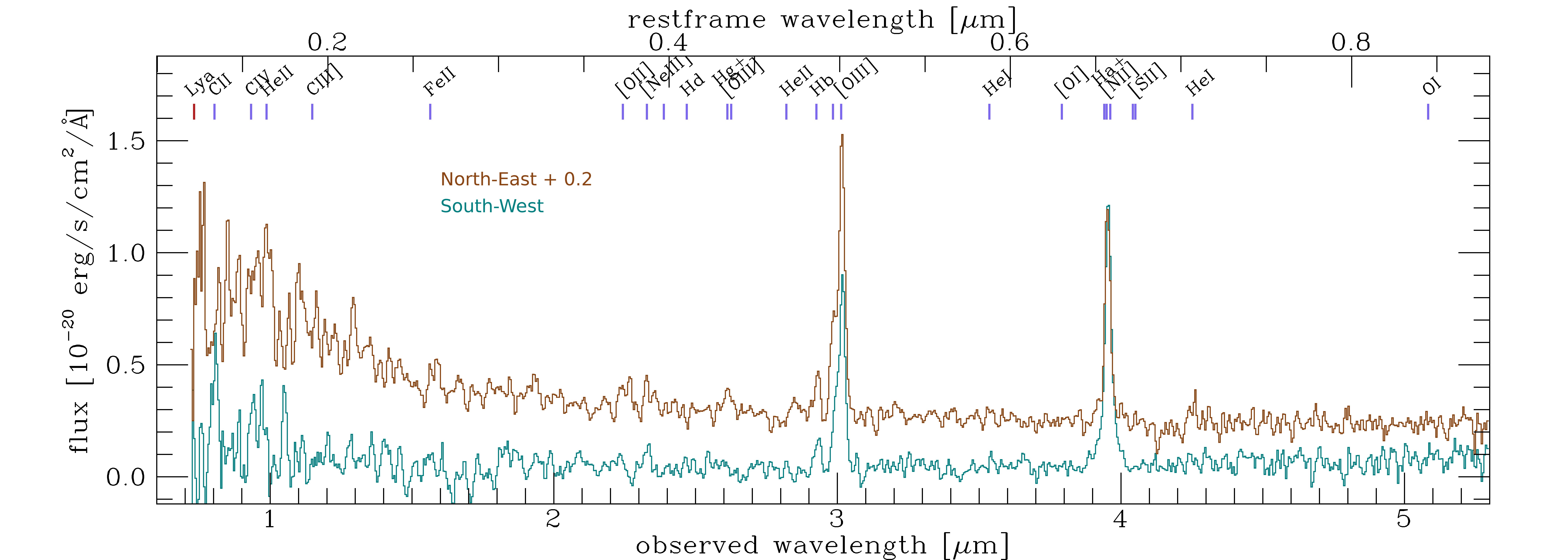}
    \caption{Two-by-two spaxel extractions from regions north-east (brown) and south-west (teal) of the centre of \tar, highlighting the different blue continuum slopes and rest-frame optical strong line ratios. These differences could be due to different excitation sources, such as the two central black holes, or could be related to the small blue source seen close to \tar\ in the NIRCam imaging.}
    \label{f:prismextr}
\end{figure*}

We also note a clear change in strong emission line ratios and blue continuum slope across \tar, seen to a lesser extent also in the high-resolution data. 
In Fig.~\ref{f:prismextr} we show two-by-two spaxel extractions north-east (brown) and south-west (teal) of the galaxy centre of the \tar\ prism data cube. Naturally, the extractions have lower $S/N$ compared to our main galaxy aperture extraction shown in Fig.~\ref{f:prismfit}, which is especially apparent in the rest-frame UV region. Despite this, the extracted regions clearly have different spectral slopes in the rest-frame UV. A change in the strong rest-frame optical emission line ratios is also evident. These differences could be explained through different excitation sources, such as two black holes, or could be the imprint of the small blue source closest to \tar\ which is seen in the NIRCam imaging.
We defer a detailed analysis of the prism data and the continuum properties of \tar\ and its potential satellites to future work.

\begin{table*}
\caption{$R$$\sim$$100$ emission line fluxes.}\label{t:fluxes_main_prism}
\begin{tabular}{@{}ccccccccc@{}}
\hline
    \civll & \heiiuv  & \ciiil & \oiill & \neiiib & \neiiir & \hg+\oiiia & H$\beta_{\rm narrow}$ & \oiiil \\ 

    $6.2\pm1.8$  & $2.4\pm1.2$  & $1.8\pm0.8$ & $1.7\pm0.2$ & $2.3\pm0.2$ & $0.5\pm0.2$ & $1.4\pm0.2$ & $4.2\pm0.5$ & $28.5\pm0.5$ \\
\hline

  He~{\sc{i}}$_{5876}$ & H$\alpha_{\rm narrow}$ & H$\alpha_{\rm BLR}$ & $[{\rm N~\sc{II}}]_{6583}$ & \siid & He~{\sc{i}}$_{7065}$ & \oi & & $\log(\rm EW_{He~II})$\\

  $1.0\pm0.2$ & $20.5\pm1.0$ & $12.9\pm1.9$ & $2.4\pm1.1$ & $0.6\pm0.1$ & $1.0\pm0.2$ & $1.0\pm0.3$ & & $1.1\pm0.5$\\

\end{tabular}
\tablefoot{
Observed $R$$\sim$$100$ emission line fluxes with $S/N\gtrsim2$ in units of $10^{-18}$~erg/s/cm$^2$ and the equivalent width of \heiiuv\ in \AA, extracted from the region outlined in orange in Fig.~\ref{f:overview}, corresponding to the fit shown in Fig.~\ref{f:prismfit}. Due to the lower resolution of the prism, the two broad \ha\ components, as well as several emission lines doublets and close pairs, are blended. Specifically, in addition to what is indicated below,   \neiiib\ is also blended with He~{\sc{i}}$_{3889}$ and \neiiir\ would be blended with H$\epsilon$, although in this case the contribution would be negligible.
}
\end{table*}

\section{Flux and kinematic maps of \oiii}\label{a:narrowmaps}

\begin{figure*}
    \centering  
    \includegraphics[width=0.3\textwidth]{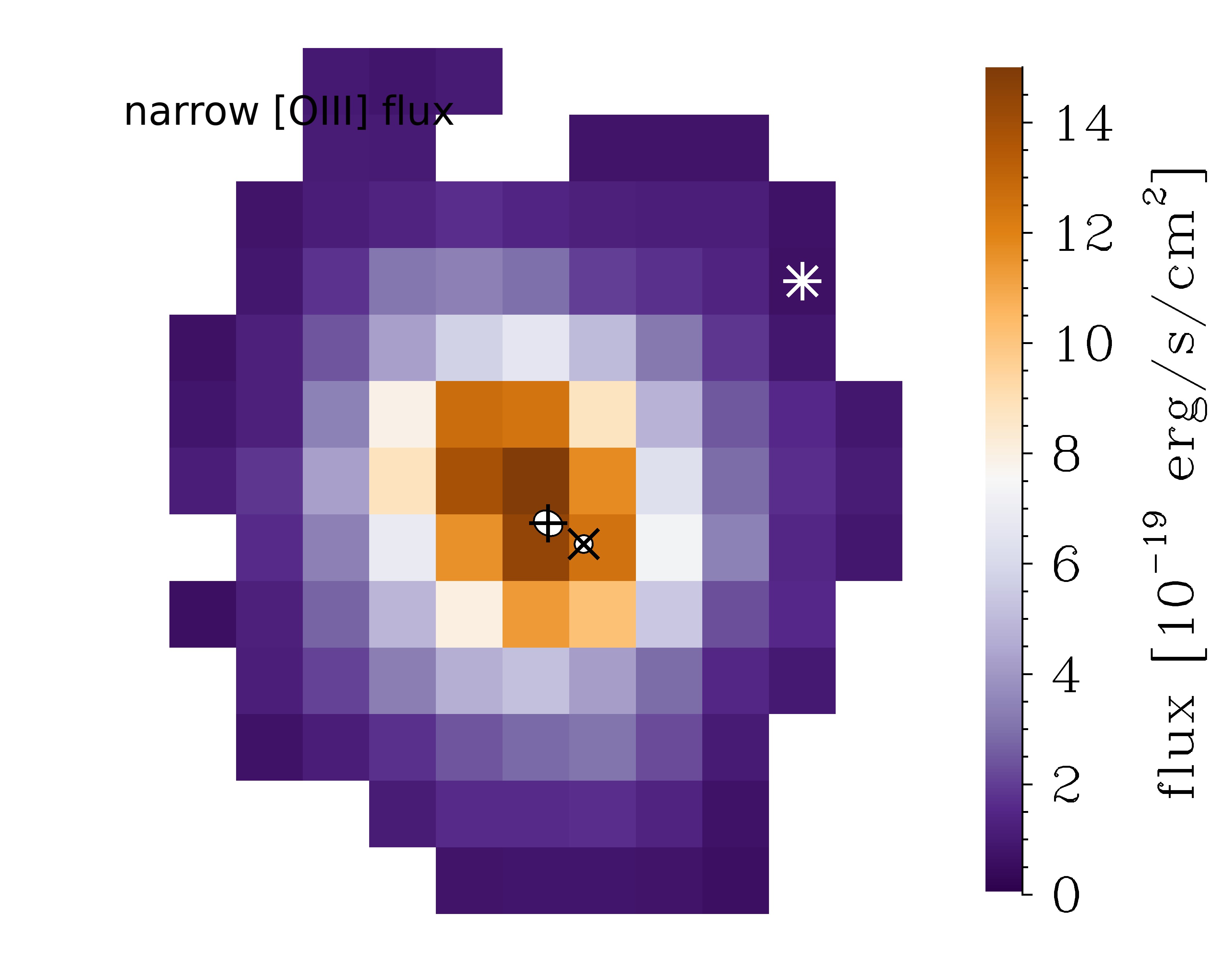}     \includegraphics[width=0.3\textwidth]{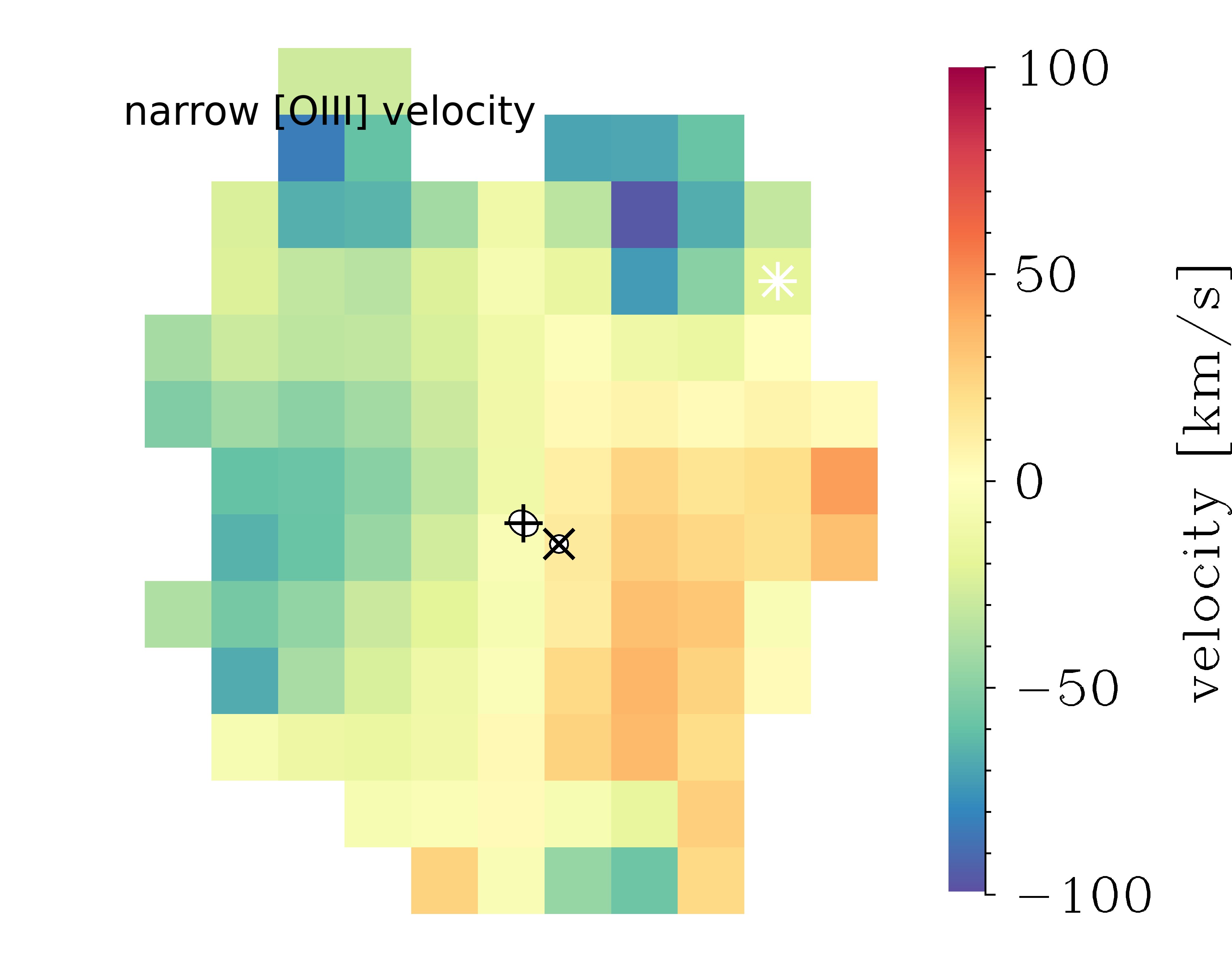}
    \includegraphics[width=0.3\textwidth]{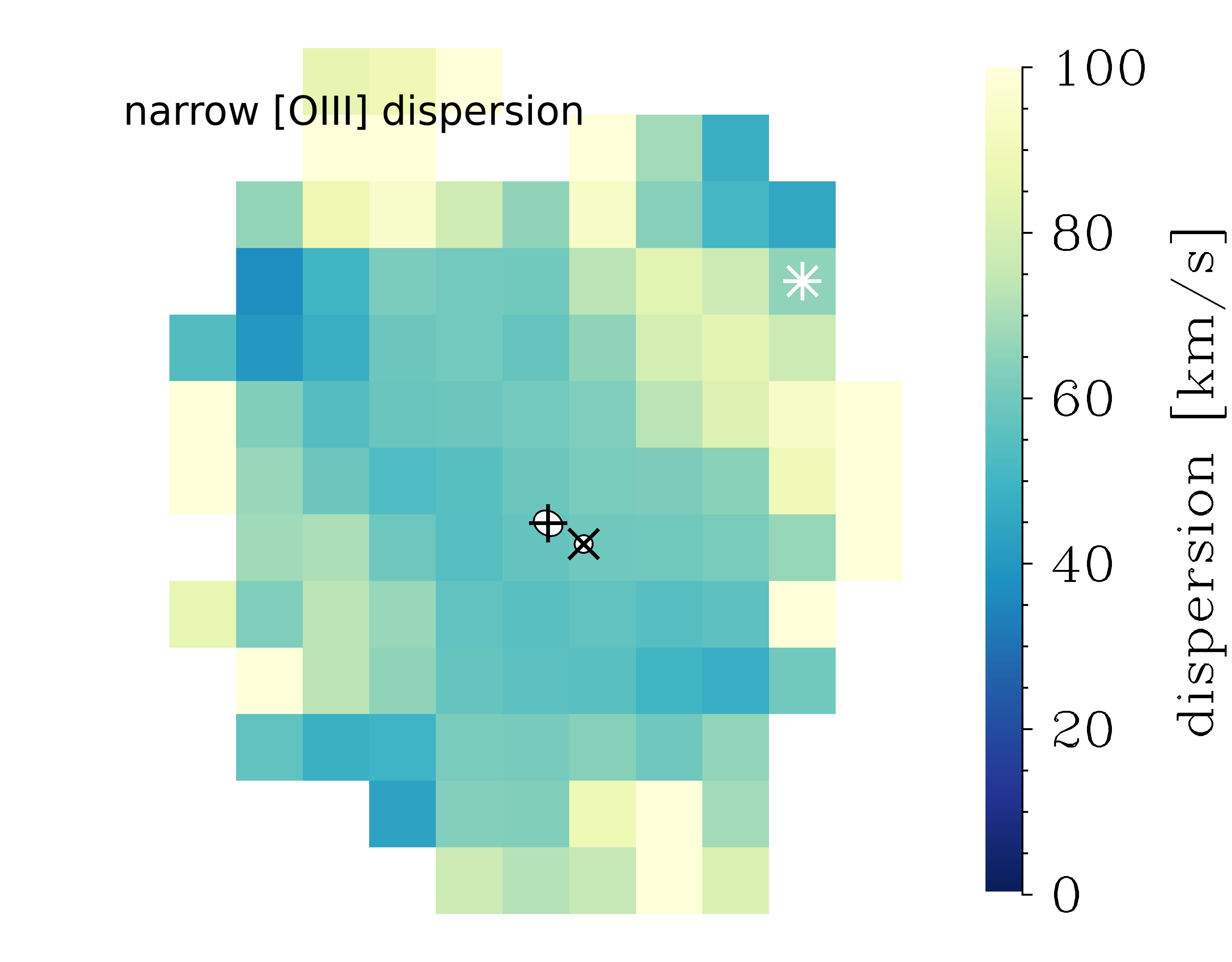}
    \caption{Maps of the \oiii\ flux, velocity, and velocity dispersion. We mark the positions of the three black holes as a cross (AGN~I), plus (AGN~II), and star (AGN~III). We observe a smooth velocity gradient of about $\Delta v_{\rm obs}\sim100$~km/s in the ENE-WSW direction (uncorrected for the PSF and inclination), and relatively constant velocity dispersion (corrected for instrumental resolution) throughout the main part of \tar.
    }
    \label{f:oiii_kinematics}
\end{figure*}

In Fig.~\ref{f:oiii_kinematics} we show maps of the (narrow) \oiiil\ flux, observed velocity and observed velocity dispersion, obtained from our spaxel-by-spaxel fitting described in Sect.~\ref{s:blrs}. We observe a fairly smooth velocity gradient of about $\Delta v_{\rm obs}\sim100$~km/s in ENE-WSW direction. 
In the NW region of \tar, we find indication for a change in velocities, as well as an increase in velocity dispersion from the galaxy centre towards AGN~III. This could support the notion that BH$_3$ is in the process of either being accreted onto \tar\ in a minor merger event or being ejected from \tar\ as a consequence of a previous black hole merger, i.e.\ in a recoil event where some of the gaseous material bound to the black hole is dragged along. 
However, we caution that the \oiii\ $S/N$ in these outer regions is fairly low, and we thus prefer not to draw strong conclusions regarding the nature of AGN~III based on these kinematic features.

\end{appendix}

\end{document}